\documentclass[prb,twocolumn,showpacs,eqsecnum,unsortedaddress]{revtex4-1}
\usepackage{graphicx}
\usepackage{amsmath,amssymb,mathrsfs}

\let\eps=\epsilon
\let\veps=\varepsilon
\let\up=\uparrow
\let\down=\downarrow

\def \calG {\mathcal{G}}
\def \beq {\begin{equation}}
\def \edq {\end{equation}}
\def \bes {\begin{subequations}}
\def \eds {\end{subequations}}
\def \beqn {\begin{equation*}}
\def \edqn {\end{equation*}}
\def \nn  {\nonumber}
\def \dag {\dagger}

\def \up {\uparrow}
\def \down {\downarrow}
\def \eps {\epsilon}
\def \sm {\sigma}
\def \bsm {\bar{\sigma}}

\def \veps {\varepsilon}
\def \calu {{\cal{U}}}
\def \calh {{\cal{H}}}

\def \calz {{\cal{Z}}}
\def \cald {{\cal{D}}}
\def \calg {{\cal{G}}}
\def \calp {{\cal{P}}}

\def \calt {{\cal{T}}}
\def \calr {{\cal{R}}}

\def \scrg {{\mathscr{G}}}

\def \wtk {\widetilde{k}}
\def \sgn {\text{sgn}}
\def \wtd {\widetilde{d}}
\def \wtc {\widetilde{c}}
\def \wV {\widetilde{V}}
\def \wteps {\widetilde{\veps}}
\def \wtU {\widetilde{U}}
\def \wtn {\widetilde{n}}

\begin{document}
\title{Transport properties of a molecule embedded in an Aharonov-Bohm
  interferometer}

\author{Jong Soo Lim}%
\affiliation{Department de F\'isica, Universitat de les Illes Balears,
  E-07122 Palma de Mallorca, Spain }%
\affiliation{Department of Physics, Korea University, Anam-Dong Seongbuk-Gu,
  Seoul 136-701, Korea}%
\author{Rosa L\'opez}%
\affiliation{Department de F\'isica, Universitat de les Illes Balears,
  E-07122 Palma de Mallorca, Spain }%
\author{Gloria Platero}%
\affiliation{Teor\'{\i}a de la Materia Condensada, Instituto de Ciencia
  de Materiales de Madrid (CSIC) Cantoblanco,28049 Madrid, Spain}
\author{Pascal Simon}%
\affiliation{Laboratoire de Physique des Solides, CNRS UMR-8502, Universit\'{e} Paris Sud, 91405 Orsay Cedex, France}
\date{\today}

\begin{abstract}
We theoretically  investigate the transport properties of
a molecule embedded in one arm of a mesoscopic Aharonov-Bohm
interferometer.  Due to the
presence of phonons the molecule level position ($\varepsilon_d$) and the
electron-electron interaction ($U$) undergo a \emph{polaronic shift} which
affects dramatically the electronic
transport through the molecular junction. When the electron-phonon
interaction is weak the linear conductance presents
Fano-line shapes as long as the direct channel between the electrodes is
opened. The observed Fano resonances in the linear conductance are originated from the interference between the
spin Kondo state and the direct path.  For  strong enough electron-phonon interaction,
the electron-electron interaction is renormalized
towards negative values, {\it i.e.} becomes effectively attractive. 
This scenario
favors fluctuations between
the empty and doubly occupied charge states and therefore promotes 
a charge Kondo effect. 
However, the direct path between the contacts breaks the
electron-hole symmetry which can efficiently suppress 
this charge Kondo effect. Nevertheless, we show that a proper tuning of the gate voltage is able to revive  the Kondo resonance.
Our results are obtained by using
the Numerical Renormalization approximation to compute the electronic
spectral function and the linear conductance.
\end{abstract}

\maketitle
\section{Introduction}
\emph{Single-Molecular transistors} are seen as basic elements to
build nanometer-scale electronic
devices.\cite{Reed97,Nit03,bookreview}~Such nanostructures
intermingle  both the electronic and mechanical degrees of freedom
and therefore give rise to new physical
scenarios.\cite{Park00,Park02,Jeon03,Saz04,Yu04,Pas05,Tsu05}~Among
their advantages, they offer a tremendous diversity and new
potential functionalities. An archetype are the single-$C_{60}$ and
-$C_{140}$ transistors in which single electron tunneling events are
used to excite and probe the quantized mechanical degrees of freedom or
\emph{vibrons} of these molecules.\cite{Park00,Yu04}~Recently these features have been also
observed in suspended carbon nanotubes.\cite{cntreferences,leturcq}~Importantly, these experiments
shows a rather strong electron-vibron coupling which is consistent with the observation of the Franck-Condon 
blockade.\cite{leturcq}  The observation of a wide
range of electron-vibron couplings in carbon nanotubes , i.e., from weak to strong
couplings may motivate new experiments to
test many of the theoretical predictions in vibron-assisted transport
through molecules.  Furthermore due to their flexibility, single molecules and especially carbon nanotubes can
be contacted to non-metallic leads such as ferromagnetic or superconducting
leads and they can even exhibit many-body physics such as the Kondo effect
when they are strongly coupled to the
reservoirs.~\cite{ferromagnetic,superconductors}

As we pointed out above the electronic motion through single molecules is sensitive to
vibrational modes that  affect dramatically the current flow.\cite{Park00,Park02,Jeon03,Saz04,Yu04,Pas05,Tsu05,Braig03,Flensberg03,Cornaglia,Mitra04,Koch05,Paaske05,Vega06,Myung07,Galperin06,Martin08}~The center-of-mass motion mode modulates
asymmetrically the coupling between the molecule and may open new channels for
transport. The observation of multiple features in the current-voltage
characteristic with identical spacing reflects the excitation of an integer
number of a well defined vibronic
mode.\cite{Park00,Park02,Jeon03,Saz04,Yu04,Pas05,Tsu05,cntreferences}~Symmetric
modes with an electron-vibron Holstein type of coupling may cause the
Franck-Condon renormalization of the tunneling between the electrodes
and the molecule. Such a mechanism suppresses  the sequential and cotunneling transport channels through
the molecule\cite{Flensberg03}.~Importantly, the electron-vibron interaction leads to the \emph{polaron shift}
in which the charging energy $U$ become effectively negative and the molecular
level ($\veps_d$) is renormalized.\cite{Hols59,Koch05}~As pointed
out in Ref.~[\onlinecite{Koch05}] this intriguing aspect modifies
the transport scenario opening a new transport channel through the
molecular transistor. Owing to \emph{negative} $U$, doubly occupied
molecular states become energetically favored. Thus, when the gate
voltage is properly adjusted so that the energy of the double
occupancy  is tuned to the energy of the empty state, the molecule
is at ``resonance'' for the tunneling of electrons in
pairs.\cite{Koch05,Myung07}~This regime is known as the \emph{strong
electron-vibron coupling regime}. In chemistry, a variety of molecules in
solution, have shown the
feature known as \emph{potential inversion} in analogy to the negative
effective charging energy in the present context.~\cite{potentialinversion} In
addition, the strong electron-vibron coupling regime can be achieved in 
vibrating carbon nanotube quantum dots as some experiments have already probed.~\cite{cntreferences}

The functionality of such devices depends greatly on the ability to preserve 
phase coherence  through
the molecule.\cite{Datta,han01,see03,key03}~ Since the phase of electron
wave function changes during
 the transport, it is of great interest to have access to the phase of the transmission amplitude
 in order to give a full characterization of the coherent transport through
 these devices. For
 this purpose we investigate the
 widely known Aharonov-Bohm (AB) effect as a valuable tool to investigate quantum coherence of electrons.  When the coherence of
a circulating electron wave packet enclosing a magnetic flux $\Phi$ is
preserved, this results in  an extra flux-dependent
phase shift. In the simplest realization of an AB interferometer, an incoming electronic wave function
splits into two paths, which join again into the outgoing electronic wave function. Applying a magnetic flux which
threads this closed geometry, the outgoing
wave function acquires a flux-dependent phase, $\varphi=e\Phi/\hbar c$, where
$\Phi=B S$ is the flux ($B$ is the
applied magnetic field, $S$ is the enclosing surface and $e$ is the electron charge). As a consequence, the transmission is a periodic
function of $\varphi$. The evolution of the electronic phase in AB geometries
 (ring and ring-like structures) was investigated in a series of pioneering works.\cite{gef84,but85a,but85b,but92}~ Their theoretical
predictions were tested in many type of
experiments.\cite{webb85,yac95,Ya96,buk98,spri00,holl01,kob02,ful01,ji00,amn02}

\begin{figure}\centering
\includegraphics*[width=85mm]{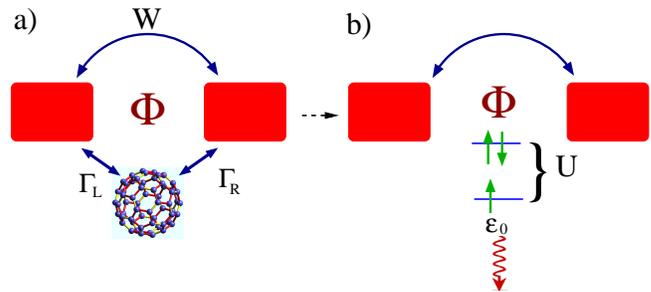}
\caption{(Color online). (a) Schematic of the molecule embedded in an
  Aharonov-Bohm interferometer. (b) Energy level diagram for the molecule.}
\label{fig:1}
\end{figure}

In our particular case, we consider a molecular state occupied by an
unpaired electron inserted in one of the arms of the AB
interferometer. The other charge states with lower energies are
doubly occupied and are therefore not active transport channels. 
The other arm acts as a direct path between the two
electrodes [see Fig.~\ref{fig:1}]. This situation corresponds to the
case of single-molecules with a non vanishing spin that behaves as
magnetic impurities\cite{Park02} exhibiting many-body physics such
as the Kondo effect.\cite{hew93}~At sufficiently low temperatures,
the localized spin residing in the molecule is able to form a
many-body spin singlet state with the delocalized electrons in the
electrodes. In this manner the molecular spin is effectively
screened and a narrow quasi-particle resonance emerges in the local
density of states (DOS) at the molecule site. The width of such a
singularity is related with the natural energy scale of this
phenomenon given by the Kondo temperature, $T_K$. Thus, $T_K$ is a
measure of the binding energy of the Kondo many-body singlet state.
In transport through quantum dots, the Kondo effect leads to the
unitary conductance as predicted theoretically\cite{theory} and
demonstrated experimentally.\cite{experiment}~The unitary limit of
the conductance in the Kondo regime is related to a transmission
phase $\delta$ equal to $\pi/2$.~\cite{hew93} This prediction has
been tested by embedding a quantum dot in one of the arm of an
Aharonov-Bohm ring.\cite{ji00,Wiel00}~Theoretical studies addressing
this problem have revealed the strong sensitivity of the phase shift
to Kondo
correlations.~\cite{Gerland00,bul01,kon01,tae01,Rej03,sil03,pascal05,Aha05}
In our particular system we expect striking and interesting
transport features due to the interplay of the electron-phonon (e-ph)
coupling and  Kondo correlations. In the
Coulomb blockade regime,
%it has been shown that depending on the
%parameter $2\lambda^2/\omega_0$ ($\lambda$ is the e-ph
%coupling, and $\omega_0$ is the phonon frequency mode)
experiments
performed with molecules such as $C_{60}$ and $C_{140}$ showed an
internal mode of vibration of $\hbar\omega_0\approx 5-10
meV$~\cite{Park00,Yu04} while the vibron mode is of order $1~meV$ in carbon nanotubes.\cite{leturcq}
This energy is comparable or even larger
than the Kondo temperature observed in these molecules which is of
the order of few Kelvin. Additionally,  for some molecules the Coulomb
charging energy becomes considerably reduced due to the effective screening
coming from the electrodes. In this way the electronic and some vibrational
energy modes (typically of the order of $0.01$-$0.1$ eV) may become comparable. This is for example the case  where the
interplay between the Kondo effect and the vibrons leads to a rich physical
phenomena reflected in the molecular transport as shown below.

This paper is organized as follows, in Sec.~\ref{model} we introduce the
model Hamiltonian. We discuss different  transport regimes
depending on the strength of the e-ph  coupling, the local mode frequency,
$\omega_0$  and the electron-electron (e-e) on-site interaction $U$. In sec.~\ref{nrg}, we
briefly present our 
theoretical approach based on the numerical renormalization group (NRG). Expressions for the
current and
linear conductance are derived in Sec.~\ref{transport}. Our numerical
results are detailed in Sec.~\ref{results}. Finally, we conclude this work by summarizing
our main findings in Sec.~\ref{conclusions}. We add two Appendices that
help  understanding the work. In
Appendix~A, we obtain the effective Hamiltonian for the different transport
regimes. Appendix~B is devoted to calculate explicitly the molecular Green function and the
phonon Green function using the equation-of-motion approach.

\section{Anderson-like and Kondo-like Hamiltonians}
\label{model} In the following we introduce the model Hamiltonian in order to
describe the molecular transport through the Aharonov-Bohm
interferometer. In a first step, we  discuss the model we use to describe a
isolated molecule. Next, we insert the molecule in the AB
interferometer and analyze how the Kondo correlations emerge.
\subsection{Isolated molecule coupled to a single local phonon mode}
In order to describe the isolated molecule in the presence of
vibrons,\cite{Hols59} we use the following model Hamiltonian
\beq  \calh_{Hol}=\calh_{M}+\calh_{e-ph}+\calh_{ph}\,,
\edq
 where
\bes
\begin{align}
\calh_{M} &= \sum_{\sigma} \veps_d d_{\sigma}^{\dag} d_{\sigma} + U n_{d\up}n_{d\down}, \\
\calh_{e-ph} &= \lambda (a+a^{\dag}) \left(n_{d} - 1\right), \\
\calh_{ph} &= \omega_0 a^{\dag}a.
\end{align}
\eds
The isolated molecule is described by the hamiltonian $\calh_{M}$:
The operator $d_{\sigma}^\dagger$ ($d_{\sigma}$) is the creation
(annihilation) operator of one electron in the unique active
molecular orbital level ($\varepsilon_{d}$) for transport.
$U$ is the \emph{on-site} Coulomb interaction, and
$n_{d\sigma}=d_{\sigma}^\dag d_{\sigma}$ is the quantum occupation of the molecular
level  per spin ($n_{d}=\sum_\sigma n_{d\sigma}$). $a^{\dag}(a)$
is the creation (annihilation) operator for an elementary quantized mechanical
excitation or vibron which we view here as a local phonon mode with a frequency $\omega_0$. 
In order to simplify our
description of the e-ph interaction we only account for single mode
for phonons corresponding to any of the vibrational normal modes of the
molecule.\cite{Braig03,Flensberg03,Cornaglia,Mitra04,Koch05,Paaske05,Vega06,Myung07,Galperin06,Martin08}~This is the extensively used \emph{Holstein} model to describe in a
simple fashion the observed molecular energy  spectrum.\cite{Hols59}~This is a
good approximation as long as the step-like features in the $I$-$V$
characteristic of some molecules are attributed to excitations of a well
defined single
mode.\cite{Park00,Park02,Jeon03,Saz04,Yu04,Pas05,Tsu05,cntreferences}
The generalization to more modes is straightforward.

\subsubsection{A first  insight on the effect of the  electron-phonon coupling}\label{sec:elph}

In order to gain some first simple  insight about the effect of the
vibrational modes on the electronic properties, it is convenient to
describe the phonon degrees of freedom by a simple harmonic
oscillator potential coupled to the electrons. This can be written
as \beq
\begin{split}
V(x) &= \frac{1}{2} m \omega_0^2 x^2 - \lambda(a^{\dag}+a)(n_d-1)
 \\
&= \frac{1}{2} \omega_0 \left(\frac{x}{\ell_{osc}}\right)^2 - \sqrt{2}\lambda (n_d-1) \frac{x}{\ell_{osc}} \\
&= \frac{1}{2}\omega_0 \left( \frac{x}{\ell_{osc}} - \frac{\sqrt{2}}{\omega_0} \lambda(n_d-1) \right)^2 - \frac{\lambda^2}{\omega_0} (n_d-1)^2\,,
\end{split}
\edq
where
\bes
\begin{align}
\ell_{osc} &= \frac{1}{\sqrt{m\omega_0}}\,, \\
x &= \frac{1}{\sqrt{2 m \omega_0}} (a^{\dag}+a) \,.
\end{align}
\eds Thus, the effects of the e-ph coupling leads to both a
shift of the average position of the harmonic oscillator and an
additional negative energy shift corresponding to the energy gain
due to polaron-formation. By rewriting this polaronic shift, we can
anticipate the resulting renormalizations:
\beq\label{effectivepotential} - \frac{\lambda^2}{\omega_0}
(n_d-1)^2 = -\frac{\lambda^2}{\omega_0}n_d(n_d-1) +
\frac{\lambda^2}{\omega_0}n_d\,. \edq Reabsorbing these terms into
the original Hamiltonian, we obtain the renormalizations
\beq\label{polaron} U_{\rm eff} \quad = \quad U -
\frac{2\lambda^2}{\omega_0}, \qquad \veps_{d,\rm eff} \quad = \quad
\veps_d + \frac{\lambda^2}{\omega_0}\,. \edq Note that the potential
minimum position now depends on the occupation number of the dot.
The renormalization of the e-e interaction gives rise to very
different physical scenarios depending on the e-ph coupling
$\lambda$. From Eq.~(\ref{polaron}) we see that $U_{\rm eff}$ can be
negative when $2\lambda^2/(U\omega_0)>1$. This corresponds to the so-called
strong e-ph coupling regime where the e-ph interaction induces an 
effective attraction between electrons. On the other hand, for a weak e-ph
coupling, i.e., when 
$2\lambda^2/(U\omega_0)<1$, the bipolaron quasiparticle is not energetically favorable since
the induced effective e-e interaction remains positive.

\subsection{Including the leads: Kondo correlations}

Let us assume that a molecule is attached to the source and drain
electrodes and placed in the lower arm of an Aharonov-Bohm
interferometer (see Fig.~\ref{fig:1}). The upper arm connects
directly the source and drain contacts by a direct path. The
simplest Hamiltonian able to describe the system is the
Anderson-Holstein Hamiltonian:
\begin{equation}
\calh = \calh_{C} + \calh_{Hol} + \calh_{T},
\label{eq:model}
\end{equation}
where
\bes
\begin{align}
\calh_{C} &= \sum_{\alpha=L/R,k,\sigma} \veps_{k} c_{\alpha k\sigma}^{\dag}c_{\alpha k\sigma}
\\ &+ \sum_{k,k',\sigma} \left(W e^{i\varphi} c_{Rk'\sigma}^{\dag}c_{Lk\sigma}+h.c.\right),\\
\calh_{T} &= \sum_{\alpha=L/R,k,\sigma} \left( V_{\alpha} c_{\alpha
  k\sigma}^{\dag} d_{\sigma} + h.c.\right),
\end{align}
\eds where we have defined $c_{L(R),k,\sigma}^\dagger$ [$c_{L(R),k,\sigma}$] is the
creation (annihilation) operator for an electron in the state $k$
with spin $\sigma$ in the lead $L(R)$. We have assumed in our
analysis that the leads are non polarized such that the dispersion
relation $\veps_k$ is spin independent. $V_{\alpha}$ is the
tunneling amplitude for the $\alpha$-electrode with the molecule.
$W$ is the amplitude probability for electron transfer from the left
electrode to the right one, and $\varphi=2\pi\Phi/\phi_0$ is the
Aharonov-Bohm phase, with $\phi_0=hc/e$ being the quantum flux.

Kondo physics arises when the molecule  is strongly contacted to the reservoirs.\cite{hew93,Park02}
~High-order tunneling processes between the delocalized
electrons in the contacts and the localized electron in the molecule lead
to screen effectively the molecular spin and to create the Kondo state. Then, a
spin singlet many-body state is formed by the conduction band electrons and
the localized unpaired  electron in the molecule. As a result, the molecular DOS shows
a very narrow peak at the Fermi energy ($E_F$).\cite{hew93} This is the so-called Kondo resonance.
In order to see how Kondo physics emerges in the presence of
e-ph interaction, let us first perform a
series of transformations of $\calh$.

The first step
%as we have already seen,
 consists in  eliminating the e-ph coupling
in $\calh_{M}$ by applying the Lang-Firsov canonical
transformation~\cite{Mahanbook,Lang63} to the full Hamiltonian $\widetilde\calh\to
\calu^{\dag} \calh\,\calu$ where the operator $\calu^{\dag}$ is defined as
\begin{equation}\label{transLF}
\calu^{\dag} = \exp\left[ \frac{\lambda}{\omega_0}(n_d-1)(a^{\dag} - a) \right]\,.
\end{equation}

The resulting Hamiltonian $\widetilde\calh=\widetilde\calh_{C}+\widetilde\calh_{M}+ \widetilde\calh_T$
 reads in the {\it parity basis}
\begin{eqnarray}
&&\widetilde\calh_{C} = \sum_{\ell=e/o,k,\sigma} \veps_{k\sigma} c_{\ell k\sm}^{\dag}c_{\ell k\sm}
\\ \nonumber
&&+ \sum_{k,k',\sigma} W \left(c_{ek'\sm}^{\dag}c_{ek\sm} -
c_{ok'\sm}^{\dag}c_{ok\sm}\right)\,,  \\ \nonumber
&&\widetilde\calh_{M}
= \omega_0 a^{\dag} a + \left(\veps_d + \frac{\lambda^2}{\omega_0}\right) n_d
+ \left(U - \frac{2\lambda^2}{\omega_0} \right) n_{d\up}n_{d\down}\,,
 \\ \nonumber
&&\widetilde\calh_T = \sqrt{2} V \sum_{\ell,k,\sm}
\Biggr[ e^{-\lambda(a^{\dag}-a)/\omega_0} \cos(\varphi/2) c_{e k\sm}^{\dag}
  d_{\sm}
\\ \nonumber
&+& e^{-\lambda(a^{\dag}-a)/\omega_0} \sin(\varphi/2) c_{o k\sm}^{\dag} d_{\sm} + h.c. \Biggr]\,,
\end{eqnarray}
where we have defined  the even-odd parity basis as
\bes
\begin{align}
c_{e k\sm} &= \frac{1}{\sqrt{2}} \left( e^{i\varphi/2}c_{L k\sm} + e^{-i\varphi/2}c_{R k\sm} \right)\,, \\
c_{o k\sm} &= \frac{i}{\sqrt{2}} \left( -e^{i\varphi/2}c_{L k\sm} + e^{-i\varphi/2}c_{R k\sm} \right)\,,
\end{align}
\label{eq:evenodd}
\eds
and $V_L = V_R \equiv V$. Notice that
now the vibronic degrees of freedom appear in a nontrivial
manner in the  tunneling part of the Hamiltonian. As we anticipated in Sec. \ref{sec:elph},
the charging energy and the molecular level position are renormalized
by the phonon frequency and the e-ph coupling as $U_{\rm eff}=U-2\lambda^2/\omega_{0}$ and
  $\veps_{d,\rm eff}=\veps_d+\lambda^2/\omega_0$. Due to this
  renormalization $U_{\rm eff}$ can be \emph{negative} inducing a
  bipolaronic attraction between the electrons (when $2\lambda^2/(\omega_0
  U)>1$). In this case and when $\Delta=\veps_d+U=0$ (that corresponds to the
  particle-hole or symmetric case), the zero and the doubly occupied
  molecule states are degenerate
  and  the low-energy
  excitations consists of charge fluctuations with a large gap for the spin
  fluctuations. These energy excitations can be described in terms of an effective
  Kondo model in the charge sector where the molecular charge fluctuates between $n_d=0$ and
  $n_d=2$. On the contrary, in the weak e-ph coupling limit(when
  $2\lambda^2/(\omega_0 U)<1$) the magnitude of the e-e
  interaction is effectively reduced but it does not favor 
the bipolaron formation. Here, the
  physics corresponds to the common spin-$1/2$ Kondo effect where the singly
  occupied molecular state is the ground state. Let us detail both regimes.

\subsubsection{The weak \emph{electron-phonon} coupling limit:  $2\lambda^2/(\omega_0 U)<1$}
To treat this limit, we first perform a second order Raileigh-Schr\"odinger perturbative
 calculation (or projection
method) in the tunneling Hamiltonian $\widetilde\calh_T$, we derive an
 effective Kondo-type Hamiltonian of the form\cite{hew93}
\beq\label{proyection1}
\calh_{eff} =  \calh_{1 0}\frac{1}{E_{1,0} - E_{0,m}}\calh_{01}
+  \calh_{12}\frac{1}{E_{1,0} - E_{2,m}}\calh_{21}\,. 
\edq 
The
eigenenergies of the isolated molecule are given in Appendix A.
%by
%\bes
%\begin{align}
%E_{1,0} &=  \veps_d + E^C\,, \\
%E_{0,m} &=  -\frac{\lambda^2}{\omega_0} + m\omega_0 + E^C\,, \\
%E_{2,m} &= -\frac{\lambda^2}{\omega_0} + 2\veps_d + U + m\omega_0 + E^C\,,
%\end{align}
%\eds
%where $E^C$ denotes the total conduction band energy.
The projectors read
\bes
\begin{align}
\calh_{01} &= \sqrt{2}\sum_{\ell,k,\sm,m} V_{\ell} \langle m | \calu^{-} | 0\rangle c_{\ell k\sm}^{\dag} d_{\sm}(1-n_{d\bsm})\,,  \\
\calh_{21} &= \sqrt{2}\sum_{\ell,k,\sm,m} V_{\ell}^{\ast} \langle m | \calu^{+} | 0\rangle \text{sgn}(\sm) d_{\sm}^{\dag} n_{d\bsm} c_{\ell k \sm} \,,
\end{align}
\eds
where we have introduced
\begin{equation}\label{upm}
\calu^{\pm} = \exp\left[\pm \lambda (a^{\dag}-a)/\omega_0 \right],
\end{equation}
and defined $\bsm = -\sm$, $V_{e} = \cos(\varphi/2)V$ and $V_{o} =
\sin(\varphi/2)V$. The explicit calculation of $\calh_{eff}$ is
given in Appendix~A. After some algebra the effective
Kondo Hamiltonian can be rewritten in a more standard form as [see Eq.~(\ref{heffective})]:
\begin{multline}\label{eq:heff}
\calh_{eff} =  \sum_{\alpha,\beta,p,q} J^{\alpha\beta} \Big[
S^z \cdot \left( c_{\alpha p\up}^{\dag}c_{\beta q\up} - c_{\alpha p\down}^{\dag}c_{\beta q\down}\right)\\
+ S^+ \cdot  c_{\alpha p\down}^{\dag}c_{\beta q\up}
+ S^- \cdot  c_{\alpha p\up}^{\dag}c_{\beta q\down}\Big] \\
+ \frac{1}{2} \sum_{\alpha,\beta,p,q,\sm} K^{\alpha\beta } c_{\alpha p\sm}^{\dag} c_{\beta q\sm}\,,
\end{multline}
where the first two lines are the Kondo interaction part while the
last line corresponds to the potential scattering part characterized by the
 coupling $K^{\alpha\beta}$. Notice that the Kondo part is
described by the \emph{isotropic} Kondo model that is characterized by a single
exchange coupling [see Eq.~(\ref{eq:jalbe}) and Eq.(\ref{eq:j2albe}) in Appendix~A]:
\begin{multline}\label{exchange}
J^{\alpha\beta}=J_\parallel^{\alpha\beta}=J_\perp^{\alpha\beta} = - 2\sum_m \Biggr(\frac{V_{\alpha}^{\ast} V_{\beta}|\langle m|\calu^{-}|0\rangle|^2}{\frac{\lambda^2}{\omega_0}+\veps_d-m\omega_0}
\\
+ \sum_{m} \frac{V_{\alpha} V_{\beta}^{\ast}|\langle m|\calu^{+}|0\rangle|^2}{\frac{\lambda^2}{\omega_0}-\veps_d-U-m\omega_0} \Biggr)\,,
\end{multline}
with
\begin{multline}
\langle m| \calu^{\pm}|0\rangle = \exp\left[-\frac{1}{2}\left( \frac{\lambda}{\omega_0}\right)^2 \right]
\frac{1}{\sqrt{m!}}\left(\pm \frac{\lambda}{\omega_0}\right)^m\,,
\\
|m\rangle = \frac{1}{\sqrt{m!}} \left(a^{\dag}\right)^m |0\rangle\,.
\label{eq:ua}
\end{multline}
To shorten the notation we omit the sub-index ph in the molecular states. From
Eq.~(\ref{upm}) one has $\calu^{-} = (-1)^m
\calu^{+}$ and for the symmetric model, $\veps_d = -U/2$, Eq.~(\ref{exchange})
becomes:
\beq
J^{\alpha\beta}(\lambda)
= \frac{8}{U} \sum_m \frac{V_{\alpha} V_{\beta}|\langle
  m|\calu^{-}|0\rangle|^2}{1-\frac{2\lambda^2}{\omega_0 U} +
  \frac{2m\omega_0}{U}}\,,
\edq
where 
\beq
J^{oo}=J\cos^2(\varphi/2)\,,\,\,\,J^{oo}=J\sin^2(\varphi/2)\,,\,\,\,J^{eo}=\frac{J}{2}\sin(\varphi)\,,
\edq
with
\beq
J=\frac{8V^2}{U} \sum_m \frac{|\langle
  m|\calu^{-}|0\rangle|^2}{1-\frac{2\lambda^2}{\omega_0 U} +
  \frac{2m\omega_0}{U}}\,.
\edq
Expanding $J^{\alpha\beta}(\lambda)$ around $\lambda=0$ we find
\beq
J^{\alpha\beta}(\lambda) \approx \frac{8}{U} V_{\alpha} V_{\beta} \left[ 1 + 2\left(1+\frac{U}{2\omega_0}\right)^{-1} \frac{\lambda^2}{U\omega_0} \right]\,.
\edq
Then, using the definition for the Kondo temperature in the isotropic Kondo
model one has 
\beq\label{eq:tkweak}
T_K \propto \exp\left[ -\frac{1}{J(\lambda)\rho_0}\right]\,.
\edq
As a result, the e-ph coupling leads to an increase of the Kondo temperature in
the weak e-ph regime as already found in
Ref.~[\onlinecite{Cornaglia}] by Cornaglia and coworkers.

\subsubsection{Strong \emph{electron-phonon} coupling: $2\lambda^2/(\omega_0 U)>1$}
In this regime  the e-ph interaction induces
a bipolaronic attraction between electrons since $U_{\rm eff}$ becomes negative. 
In this case the ground state is a molecular doublet state composed of
the doubly occupied molecular state
$|2,0\rangle$ and the empty molecular state $|0,0\rangle$ which are
degenerate. The low-energy excitations
thus consist of charge fluctuations and can be described by an effective
Kondo model in which the role of spin is played now by a pseudospin
variable that represents the two states of the doublet. Importantly,
there is no rotational invariance in pseudospin space and therefore
the effective model corresponds to an \emph{anisotropic} Kondo model with
different couplings, namely $J_{\parallel}$ and $J_{\perp}$. In order to gain
some insights about the physical properties of this regime, we employ again
the projection method. Before this, it is convenient to map the
negative-$U$ Anderson-Holtstein model to an \emph{equivalent} model with
positive interaction. This can be carried out by using a particular
particle-hole (p-h) transformation.\cite{zawa72,coleman91}~First, we
introduce new operators $\wtd$, $\wtc$ via
\begin{eqnarray}
& d_{\down} \equiv -\wtd_{\down}^{\dag},\,\,\, c_{Lk\down} &\equiv
  \wtc_{2\wtk\down}^{\dag},\,\,\,  c_{Rk\down} \equiv
  \wtc_{1\wtk\down}^{\dag}\,, \\  \nonumber
&d_{\up} \equiv \wtd_{\up},\,\,\, c_{Lk\up} &\equiv \wtc_{1k\up} ,\,\,\,c_{Rk\up} \equiv \wtc_{2k\up}\,.
\end{eqnarray}
%Under this transformation, Eq.~\eqref{eq:model} becomes
%\bes
%\begin{align}
%\calh_C &= \sum_{\ell=1/2,k,\sm} \veps_{k\sm} \wtc_{\ell k\sm}^{\dag} \wtc_{\ell k\sm}\,
%\\
%&+ \sum_{k,k',\sm} \left[\sgn(\sm) W e^{i\varphi} \wtc_{2k'\sm}^{\dag} \wtc_{1k\sm} + h.c. \right]\,,\\
%\calh_M &= \sum_{\sm} \wteps_{d\sm} \wtd_{\sm}^{\dag} \wtd_{\sm} + \wtU \wtn_{d\up}\wtn_{d\down}\,,
% \\
%\calh_T &= \sum_{\ell=1/2,k,\sm} \left( \wV_{\ell k\sm} \wtc_{\ell k\sm}^{\dag} \wtd_{\sm} + h.c.\right)\,, \\
%\calh_{e-ph} &= \lambda (a+a^{\dag})\left(\wtn_{d\up} - \wtn_{d\down}\right) \,\\
%\calh_{ph} &= \omega_o a^{\dag}a\,,
%\end{align}
%\label{eq:stmodel}
%\eds
The p-h transformation yields $\veps_{d\sm}\rightarrow\wteps_{d\sm}
= U/2 + \sgn(\sm) \Delta /2$ and $U\rightarrow\wtU = - U$ where
$\Delta = 2\veps_d + U$ and $\veps_{k\sm} = -\veps_{\wtk \sm}$ assuming the symmetry condition for the conduction band. The tunneling amplitudes are given by
\begin{eqnarray}
&\wV_{1k\up} = V_{Lk\up}, \,\, \wV_{1k\down} = V_{R\wtk\down}^{\ast}, \\
&\wV_{2k\up} = V_{Rk\up}, \,\,\wV_{2k\down} = V_{L\wtk\down}^{\ast}\,.
\end{eqnarray}
The p-h transformation maps the electronic states of the molecule to
the ones of the original model in this manner
\beq
|0\rangle \mapsto |\widetilde{\down}\rangle, \quad
|2\rangle \mapsto |\widetilde{\up}\rangle, \quad
|\down\rangle \mapsto |\widetilde{0}\rangle, \quad
|\up\rangle \mapsto |\widetilde{2}\rangle\,.
\edq
Now we perform the polaronic transformation~\cite{Mahanbook,Lang63} combined with a canonical
transformation such that the full Hamiltonian in the
parity basis [see Eq.~\eqref{eq:evenodd}] reads (see Appendix A)
\begin{multline}
\tilde\calh_C = \sum_{\ell=e/o,k,\sm} \veps_{k\sm} \wtc_{\ell k\sm}^{\dag}\,,
\wtc_{\ell k\sm} \\
+ \sum_{k,k',\sm} \sgn(\sm) W \left(\wtc_{ek'\sm}^{\dag} \wtc_{ek\sm} - \wtc_{ok'\sm}^{\dag} \wtc_{ok\sm} \right)\,,\\
\tilde\calh_M = \sum_{\sm} \left(\wteps_{d\sm}-\frac{\lambda^2}{\omega_0}\right) \wtd_{\sm}^{\dag} \wtd_{\sm} + \left(\wtU + \frac{2\lambda^2}{\omega_0}\right) \wtn_{d\up}\wtn_{d\down} \\
+ \omega_0 a^{\dag}a \,,
\\
\tilde\calh_T = \sqrt{2} \sum_{\ell=e/o,k,\sm} \Big( \wV_{\sm} e^{-\sgn(\sm)\lambda/\omega_0(a^{\dag}-a)} \cos(\varphi/2) \wtc_{e k\sm}^{\dag} \wtd_{\sm} \,
\\
+ \wV_{\sm} e^{-\sgn(\sm)\lambda/\omega_0(a^{\dag}-a)} \sin(\varphi/2) \wtc_{o k\sm}^{\dag} \wtd_{\sm} + h.c.\Big)\,,
\end{multline}
where we assumed $\wV_{1/2 \sm} = \wV_{\sm}$. The final step consists in
obtaining an effective Hamiltonian by using again the projection method. We employ
 Eq.~(\ref{proyection1}) and we refer to Appendix A for details
%Here,
% the eigenenergies and projectors $\calh_{01}$ and
% $\calh_{21}$ read (see  Apendix~A for details)
\bes
\begin{align}
%E_{1,0} &=  \wteps_d + E^C\,, \\
%E_{0,m} &=  \frac{\lambda^2}{\omega_0} + m\omega_0 + E^C\,, \\
%E_{2,m} &= \frac{\lambda^2}{\omega_0} + 2\wteps_d + U + m\omega_0 + E^C\,, \\
\calh_{01} &= \sqrt{2}\sum_{\ell,k,\sm,m} \wV_{\ell k\sm} \langle m | \calu^{-\sgn(\sm)} | 0\rangle \wtc_{\ell k\sm}^{\dag} \wtd_{\sm}(1-\wtn_{d\bsm})\,,  \\
\calh_{21} &= \sqrt{2}\sum_{\ell,k,\sm,m} \wV_{\ell k\sm}^{\ast} \langle m | \calu^{+\sgn(\sm)} | 0\rangle \text{sgn}(\sm) \wtd_{\sm}^{\dag} \wtn_{d\bsm} \wtc_{\ell k \sm} \,,
\end{align}
\eds
with $\bsm = -\sm$, and $\wV_{e\sm} = \cos(\varphi/2)\wV_{\sm}$ and
$\wV_{o\sm} = \sin(\varphi/2)\wV_{\sm}$. After the projection, the effective
Hamiltonian $\calh_{eff}$ corresponds to the \emph{anisotropic} Kondo model
with the following exchange couplings (see Appendix~A)
\bes
\begin{align}
&J_{\parallel}^{\alpha\beta} = \frac{8}{U} \sum_m \frac{V_{\alpha} V_{\beta}|\langle m|\calu^{-}|0\rangle|^2}{\frac{2\lambda^2}{\omega_0 U}-1+\frac{2m\omega_0}{U}}\,, \\
&J_{\perp}^{\alpha\beta} = \frac{8}{U} \sum_m
\frac{V_{\alpha} V_{\beta} \langle 0|\calu^{-}|m\rangle\langle m|\calu^{-}|0\rangle}{\frac{2\lambda^2}{\omega_0 U}-1+\frac{2m\omega_0}{U}}
\\
&= \frac{8}{U} \sum_m (-1)^m
\frac{V_{\alpha} V_{\beta} |\langle m|\calu^{-}|0\rangle|^2}{\frac{2\lambda^2}{\omega_0 U}-1+\frac{2m\omega_0}{U}}\,.
\end{align}
\eds 
The Kondo temperature in this case is given by\cite{Tsvelick83}
\beq\label{tkstrongdef}
T_K \propto D \left(\frac{J_{\perp}}{J_{\parallel}}
\right)^{1/J_{\parallel}\rho_0}\,. 
\edq In the limit of large $\lambda/\omega_0^2$, $J_{\perp}/J_{\parallel}$
behave as 
\beq
J_{\perp}/J_{\parallel} \approx \exp\left[-2(\lambda/\omega_0)^2
\right]\,. 
\edq 
Then, the Kondo temperature $T_K$ reads\cite{hew93} 
\beq
\label{eq:tkstrong} 
T_K \propto \exp\left[
-\left(\frac{\lambda}{\omega_0}\right)^4 \frac{\pi\omega_0}{\Gamma}
\right]\,. 
\edq In contrast to the weak e-ph coupling case, the
Kondo temperature decreases sharply with increasing $\lambda$
($\Gamma = 2 \pi \rho_0 V^2$, with $\rho_0=\rho_L=\rho_R$ being the
DOS of the conduction band electrons, we took the same for both
contacts). Equations (\ref{eq:tkweak}) and (\ref{eq:tkstrong}) are
the main results of this section.

\section{The numerical renormalization group approach}\label{nrg}

%As we have already discussed, there exists two different regimes for the
%e-ph interaction, the weak e-ph coupling regime that corresponds
%to the spin Kondo effect [$2\lambda^2/(\omega_0 U)<1]$ and the strong e-ph
%coupling regime resulting into the charge Kondo effect [$2\lambda^2/(\omega_0 U)>1]$.
In order to describe both the spin Kondo effect, occurring
for $2\lambda^2/(\omega_0 U)<1$,
and the charge Kondo effect, 
occurring for $2\lambda^2/(\omega_0 U)>1$, on the same footing,
%an unified description of both regimes we need a very powerful theoretical
%approach.
we have therefore used the Numerical Renormalization group (NRG)
technique.\cite{Wilson75,Krishnamurthy80}~Unlike other approaches, the NRG is
not restricted for the values of the parameters ($\lambda$,
$\omega_0$, $U$, $\Gamma$, $W$, $\varphi$, etc).
%Other approaches are more limited, like for
%instance, the equation of motion technique that can only be applied safely for $T>T_K$.

NRG method is employed to solve the full Hamiltonian. This allows us to compute the linear transport
properties through the AB interferometer such as the linear conductance. 
%In order to obtain high-quality quantitative predictions in the Anderson
%model, we need to implement new truncation schemes used in the NRG calculation. 
It is known that NRG method suffers from loss of accuracy coming from the
discretization of the conduction band and the truncation of high energy
states. This problem is serious in our system because the Hilbert space grows
exponentially with the number of conduction channels in the leads. Any careless
truncation can lead to significant deterioration of thr results. We thus exploit
recent improvements in the NRG procedure as well as a symmetry property of the
system to obtain high-quality predictions.\cite{nrgapproach} At very low frequencies the NRG schemes
 developed by R.~\v{Z}itko and T.~Pruschke in Ref.~[\onlinecite{rok}] and W. Hofstetter and
 G. Zarand in Ref.~[\onlinecite{nrgapproach}] yield similar results. These
 results are in agreement with those obtained by employing directly the
 Friedel-Langreth sum rule~\cite{hew93} valid only at zero temperature. However, for the
 high frequency spectrum it seems more convenient to employ the NRG code, for instance if we were interested in the vibronic sidebands, developed in
Ref.~[\onlinecite{rok}]. Since we are interested in the linear transport at
zero temperature, then
we simply use the Friedel sum rule that agrees with very refined NRG
schemes in the low frequency regime.

According to the standard NRG procedure, Eq.~\eqref{eq:model} can be rewritten
as (note that the conduction band is now discrete with states denoted by the $f$-operators):
\begin{multline}
\calh = \calh_{M} + \calh_{e-ph} + \calh_{ph}
\\ + \sum_{\sigma} \sqrt{\frac{4\Gamma}{\pi}} \left( \cos(\varphi/2) d_{\sigma}^{\dag}f_{0e\sigma} + \sin(\varphi/2) d_{\sigma}^{\dag}f_{0o\sigma} + h.c. \right) \\
+ \frac{1+\Lambda^{-1}}{2} \sum_{\alpha=e/o}\sum_{n=0}^{\infty}\sum_{\sigma} \Lambda^{-n/2} \zeta_n \left( f_{n\alpha\sigma}^{\dag}f_{n+1\alpha\sigma} + h.c.\right)
\\+ \frac{2}{\pi} \sqrt{\xi}\sum_{\sigma} \left(f_{0e\sigma}^{\dag}f_{0e\sigma} - f_{0o\sigma}^{\dag}f_{0o\sigma} \right)\,,
\end{multline}
%\begin{multline}
%\calh = \calh_{M} + \calh_{e-ph} + \calh_{ph}
%\\ + \sum_{\alpha=L/R}\sum_{\sigma} \sqrt{\frac{4\Gamma}{\pi}} \left( d_{\sigma}^{\dag}f_{0\alpha\sigma}+h.c.\right) \\
%+ \frac{1+\Lambda^{-1}}{2} \sum_{\alpha=L/R}\sum_{n=0}^{\infty}\sum_{\sigma} \Lambda^{-n/2} \zeta_n \left( f_{n\alpha\sigma}^{\dag}f_{n+1\alpha\sigma} + h.c.\right)\,,
%\\+ \frac{2}{\pi} \sqrt{\xi}\sum_{\sigma} \left(e^{i\varphi} f_{0R\sigma}^{\dag}f_{0L\sigma} + h.c. \right)\,,
%\end{multline}
with $\xi \equiv \pi^2 W^2 \rho_0^2$ and 
\bes
\begin{align}
f_{ne\sigma} &= \frac{1}{\sqrt{2}} \left( e^{i\varphi/2}f_{nL\sigma} + e^{-i\varphi/2}f_{nR\sigma} \right)\,, \\
f_{no\sigma} &= \frac{-i}{\sqrt{2}} \left( -e^{i\varphi/2}f_{nL\sigma} + e^{-i\varphi/2}f_{nR\sigma} \right)\,,
\end{align}
\eds
the parity basis. Now we apply the NRG method to solve the Hamiltonian. We define a sequence of Hamiltonians $\bar{\calh}_N$:
\begin{multline}
\bar{\calh}_N = \Lambda^{(N-1)/2} \Biggr\{ \calh_{M} + \calh_{e-ph} + \calh_{ph}
\\
+ \sum_{\sigma} \sqrt{\frac{4\Gamma_{\sigma}}{\pi}} \left( \cos(\varphi/2)
d_{\sigma}^{\dag}f_{0e\sigma} + \sin(\varphi/2) d_{\sigma}^{\dag}f_{0o\sigma}
+ h.c. \right)
\\
+ \frac{1+\Lambda^{-1}}{2}\sum_{\alpha=e/o}\sum_{n=0}^{N-1}\sum_{\sigma} \Lambda^{-n/2} \zeta_n \left(f_{n\alpha\sigma}^{\dag}f_{n+1\alpha\sigma} + h.c.\right)
\\
+ \frac{2}{\pi} \sqrt{\xi} \sum_{\sigma} \left(f_{0e\sigma}^{\dag}f_{0e\sigma} - f_{0o\sigma}^{\dag}f_{0o\sigma} \right)\Biggr\}\,,
\end{multline}
that follows the recursion relation
\beq
\widetilde{\calh}_{N+1} = \sqrt{\Lambda} \widetilde{\calh}_{N}
+ \sum_{\alpha=e/o}\sum_{\sigma} \zeta_n \left( f_{N\alpha\sigma}^{\dag}f_{N+1\alpha\sigma} + h.c.\right)\,,
\edq
with
\beq
\widetilde{\calh}_{N} = \frac{2}{1+\Lambda^{-1}}\bar{\calh}_{N}\,.
\edq
From consecutive diagonalizations of the Hamiltonian, we obtain the eigenvalues and
eigenstates. With them we build the molecular and phonon Green functions that
are needed for the calculation of the linear transport properties.
In order to improve the quality of these Green functions, we follow Jeon {\it et al.,}\cite{Jeon03} and relate the molecule and local phonon Green
functions  by using the equation of motion (EOM) technique. We relegate the
details of such a calculation to the Appendix~\ref{app2}.
\section{Linear Transport: conductance}\label{transport}
The electrical current through the left barrier is calculated through the
simple relation
\beq
I_L=-e \dot{N}_L=\frac{-i e}{\hbar} [\calh,N_L]\,,
\edq
where $N_L=\sum_{k\sigma} c_{Lk\sigma}^\dagger c_{Lk\sigma}$. After some
algebra  the current through the AB
interferometer takes the usual
Landauer-B\"uttiker form:\cite{bul01,hof01}
\beq
I=\frac{e}{h} \sum_{\sigma} \int d\eps~ \calt_{\sigma}(\eps) [f_L(\eps) - f_R(\eps)]\,,
\label{eq:current}
\edq
where $f_{L(R)}$ is the left (right) Fermi function and the transmission
probability reads
\begin{multline}
\calt_{\sigma}(\eps) = \calt_b + \sqrt{\alpha\calt_b \calr_b} \cos(\varphi) \widetilde{\Gamma} \Re\left[\calg_{d\sigma,d\sigma}^r(\eps)\right]
\\- \frac{1}{2} \left\{\alpha\left[1-\calt_b \cos^2(\varphi)\right] - \calt_b\right\} \widetilde{\Gamma} \Im \left[\calg_{d\sigma,d\sigma}^r(\eps)\right]\,,
\end{multline}
where, \beq \calt_b = \frac{4\xi}{(1+\xi)^2}\,, \edq is the
background transmission with $\xi = \pi^2 W^2 \rho_0^2$. The symbols
$\Im$ and $\Re$ denote the imaginary and real part of a complex
number and $\calg_{d\sigma,d\sigma}^r$ is the quasi-particle
retarded Green function for the localized electrons at the molecule
site. The background reflection is obtained immediately by $\calr_b
= 1 - \calt_b$. It is also useful to define  $\widetilde{\Gamma}
=\Gamma/(1+\xi)$, $\alpha = 4\Gamma_L\Gamma_R/\Gamma^2$ with $\Gamma
= \Gamma_L + \Gamma_R$ ($\Gamma_{L(R)}=\pi\rho_{L(R)}V_{L(R)}^2$, below we discuss only results for the case of
symmetric tunneling rates, i.e.,  $\Gamma_L=\Gamma_R=\Gamma=\pi\rho_0 V^2 $ with $\alpha=1$). In the linear response regime,
Eq.~\eqref{eq:current} can be written as 
\beq I = \lim_{V\to 0}
\frac{e^2 V}{h} \sum_{\sigma} \int d\eps~ \calt_{\sigma}(\eps)
f_{eq}'(\eps)\,, 
\edq 
where $f_{eq}$ is the equilibrium Fermi
function. At $k_BT=0$, 
\beq I = -\lim_{V\to 0} \frac{e^2 V}{h}
\sum_{\sigma} \int d\eps~ \calt_{\sigma}(\eps) \delta(E_F-\eps)\,.
\edq 
Thus, 
\beq G = \frac{e^2}{h} \sum_{\sigma} \calt_{\sigma}(E_F),
\quad \text{at $k_BT=0$}\,. 
\edq 
This can be further simplified by
taking the quasi-particle Green function for the localized electrons
written in a Dyson type equation as follows 
\beq
\calg_{d\sigma,d\sigma}^r(\eps) = \frac{1}{\eps - \veps_d -
\Sigma(\eps)}\,. 
\edq 
At $k_BT=0$, in the pure Kondo regime, the
imaginary part of the interaction self-energy evaluated at the Fermi
energy vanishes, then we have 
\beq \Im\left[ \Sigma(E_F) \right] = -
\frac{\widetilde{\Gamma}}{2}, \quad q = \frac{2}{\widetilde{\Gamma}}
\left\{ \veps_d + \Re\left[\Sigma(E_F)\right] \right \} \,. 
\edq
Using this notation we can write a more transparent expression for the
transmission probability in a 
generalized Fano form as: 
\beq \calt(E_F) = \calt_b \frac{(q+\mathcal{F})^2}{e^2
+1} + \alpha \frac{\sin^2(\varphi)}{q^2+1}\,, 
\edq with a Fano parameter given by 
\beq 
\mathcal{F} = -\cot(\delta_{non}) = -\sqrt{\frac{\alpha
\calr_b}{\calt_b}} \cos(\varphi)\,.
\label{eq:fano} 
\edq
\begin{figure}
\centering
\includegraphics[width=0.5\textwidth]{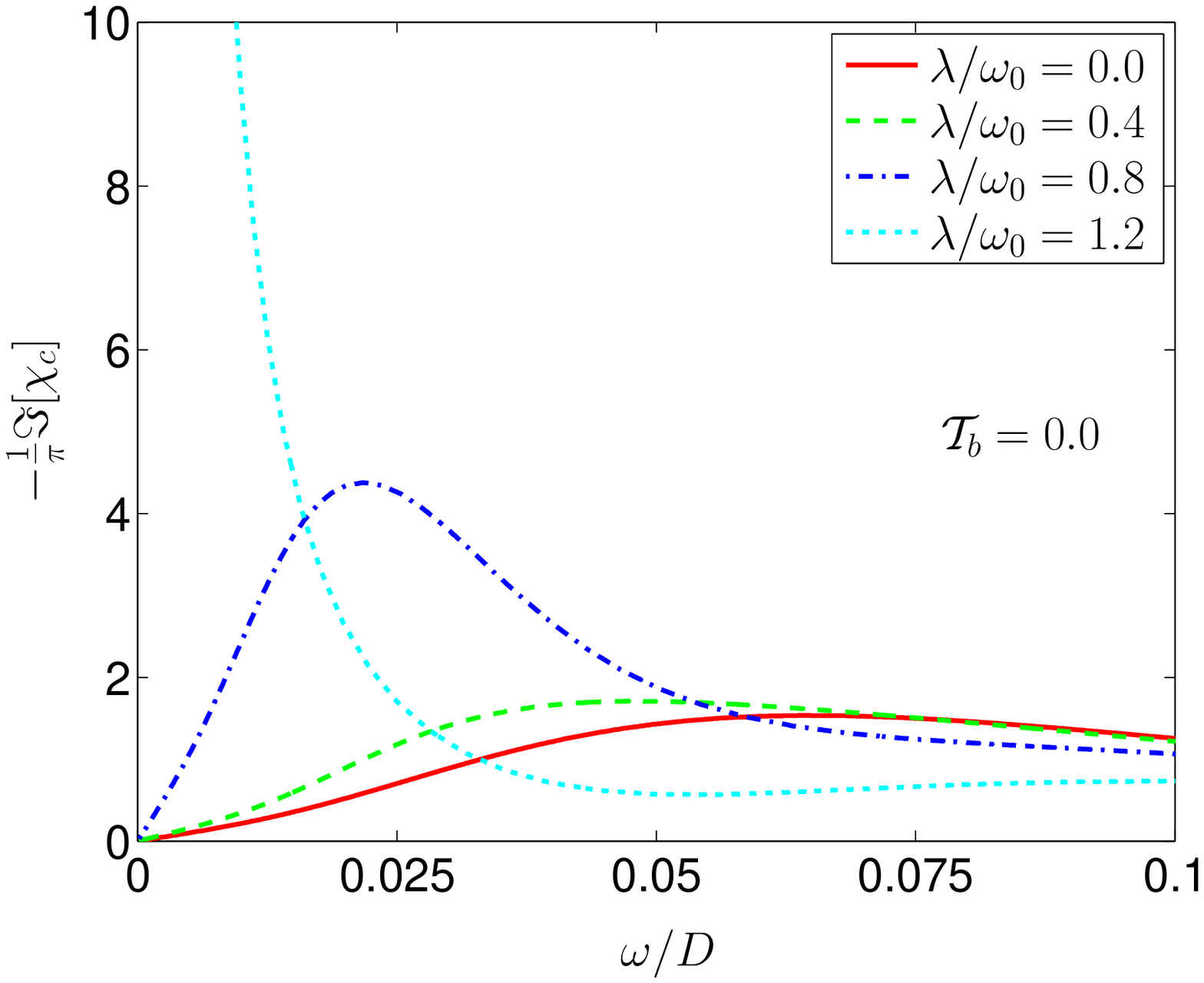}
\caption[Charge susceptibility $\chi_c(\omega)$ as a function of the $e$-$ph$
coupling $\lambda/\omega_0$]
{(Color online) Charge susceptibility $\chi_
c(\omega)$ as a function of the $e$-$ph$ coupling
$\lambda/\omega_0$. Parameters are $\veps_d=0.05$, $U=0.1$, $2\Gamma=0.0016$,
 $\varphi=0$, $\calt_b=0$, $k_BT=0$.}
\label{fig:chargesus}
\end{figure}
Notice that we have introduced $\delta_{non}$ as a non-resonant phase shift
owing to scattering of electrons through the direct path.  Alternatively, in
terms of resonant and non-resonant phase shifts the transmission probability can be written as:
\begin{multline}
\calt(E_F) = \left[ \calt_b + \alpha\calr_b\cos^2(\varphi)\right]
\sin^2(\delta_{res}-\delta_{non}) \\
+ \alpha \sin^2(\varphi) \sin^2(\delta_{res})= \calt_b
\frac{\sin^2(\delta_{res}-\delta_{non})}{\sin^2(\delta_{non})}
\\
+ \alpha \sin^2(\varphi) \sin^2(\delta_{res})\,,
\label{eq:tef}
\end{multline}
where the resonant scattering
phase shift $\delta_{res}$ can be related to the
quantum average occupation number $n_d=\sum_{\sigma} \langle d_{\sigma}^{\dag}d_{\sigma}\rangle$
using the Friedel-Langreth sum rule:~\cite{hew93}
\beq
\delta_{res} = \frac{\pi n_d }{2}={\rm tan}^{-1}\left\{\frac{\tilde\Gamma}{\veps_d+\Re\left[\Sigma(E_F)\right]}\right\}\,.
\edq
%with $n_d=\sum_{\sigma} \langle d_{\sigma}^{\dag}d_{\sigma}\rangle$ the
%quantum average occupation number.  

\section{Numerical results}\label{results}

In the following we present our numerical results based on the NRG method. We
fix $U=0.1$, $2\Gamma = 0.016$, $\omega_0 = 0.05$ and $k_BT=0$ and we vary the rest of
parameters.
We have chosen a value of the phonon vibrational frequency $\omega_0$ of the
order of the Coulomb energy scale which corresponds with the experimental
situation. Using these parameters the bipolaronic attraction occurs for $\lambda/\omega_0>1$ whereas the weak e-ph coupling limit takes
place for $\lambda/\omega_0<1$.

We start investigating how are affected the Kondo correlations by
the presence of phonons ($\lambda\neq 0$). In order to do so we expand Eq.~\eqref{eq:tef} around
the symmetric point, $\veps_d^{\star}=-U/2$ (for $\calt_b = 0$ and $\varphi=0$), we have
\beq
G(\veps_d) \approx G(\veps_d^{\star}) - G_0 \left( \frac{\pi \chi_c}{2}
\right)^2 (\veps_d - \veps_d^{\star})^2,
\edq
where, $\chi_c =\langle \langle n_d(t), n_d(0) \rangle\rangle = -i\Theta(t) \langle [n_d(t),n_d(0)]_- \rangle$ denotes the charge susceptibility.
Figure~\ref{fig:chargesus} shows the charge susceptibility $\chi_c$ as a
function of the e-ph coupling $\lambda/\omega_0$.  We see that the
charge susceptibility presents a wide peak for $\lambda=0$ (almost flat charge susceptibility
spectrum). Then, the broad peak decreases and moves towards higher frequencies
whereas a low energy peak emerges in the
spectrum as long as $\lambda/\omega_0$ grows. The low energy peak corresponds to charge
Kondo correlations that have a different energy scale ($T_K$)
than the spin Kondo correlations (corresponding to the broad peak). This
result demonstrates that the e-ph coupling induces Kondo
correlations in the charge sector reflected in the appearance of a low energy peak in $\chi_c$.

\begin{figure}
\centering
\includegraphics[width=0.45\textwidth]{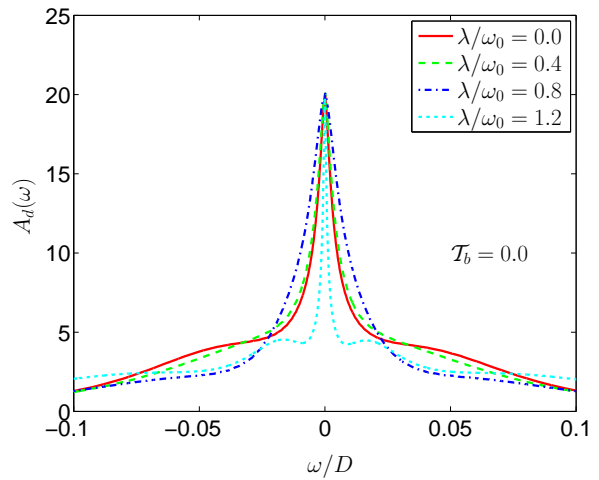}
\caption
%[Electronic spectral density $A_d(\omega)$ as a function of $\lambda/\omega_0$ and $\calt_b$]
{(Color online) Electronic spectral density $A_d(\omega)$ for different values of $\lambda/\omega_0$ and $\calt_b=0$. Parameters are $\veps_d = -0.05$, $U=0.1$, $2\Gamma=0.016$, $\varphi = 0$, and $k_BT=0$.}
\label{dos1}
\end{figure}
\begin{figure}
\centering
\includegraphics[width=0.45\textwidth]{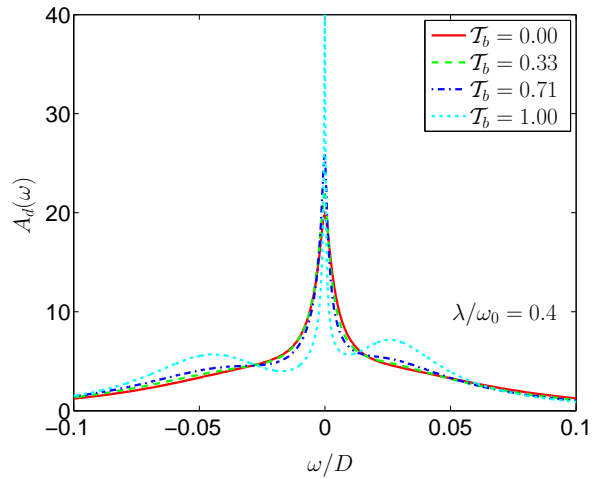}
\caption{(Color online) Electronic spectral density $A_d(\omega)$ for $\lambda/\omega_0=0.4$ and different values of
  $\calt_b$. This corresponds to the \emph{weak e-ph coupling regime}. Parameters are $\veps_d = -0.05$, $U=0.1$, $2\Gamma=0.016$, $\varphi = 0$, and $k_BT=0$.}
\label{dos2}
\end{figure}

\begin{figure}
\centering
\includegraphics[width=0.45\textwidth]{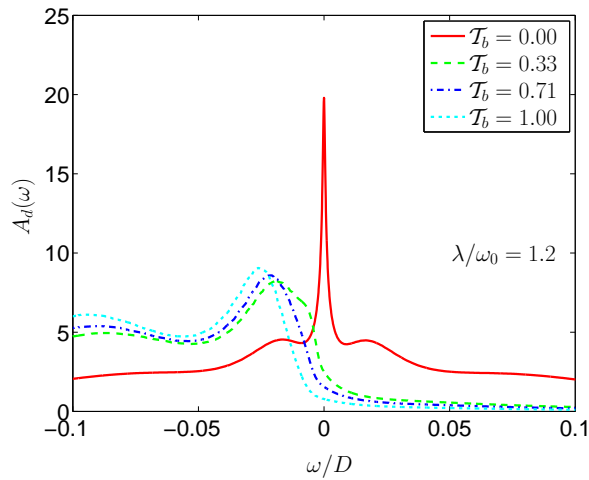}
\caption{(Color online) Electronic spectral density $A_d(\omega)$ for $\lambda/\omega_0=1.2$ and different values of
  $\calt_b$. This corresponds to the \emph{strong e-ph coupling regime}. Parameters are $\veps_d = -0.05$, $U=0.1$, $2\Gamma=0.016$, $\varphi = 0$, and $k_BT=0$.}
\label{dos3}
\end{figure}

\begin{figure}
\centering
\includegraphics[width=0.45\textwidth]{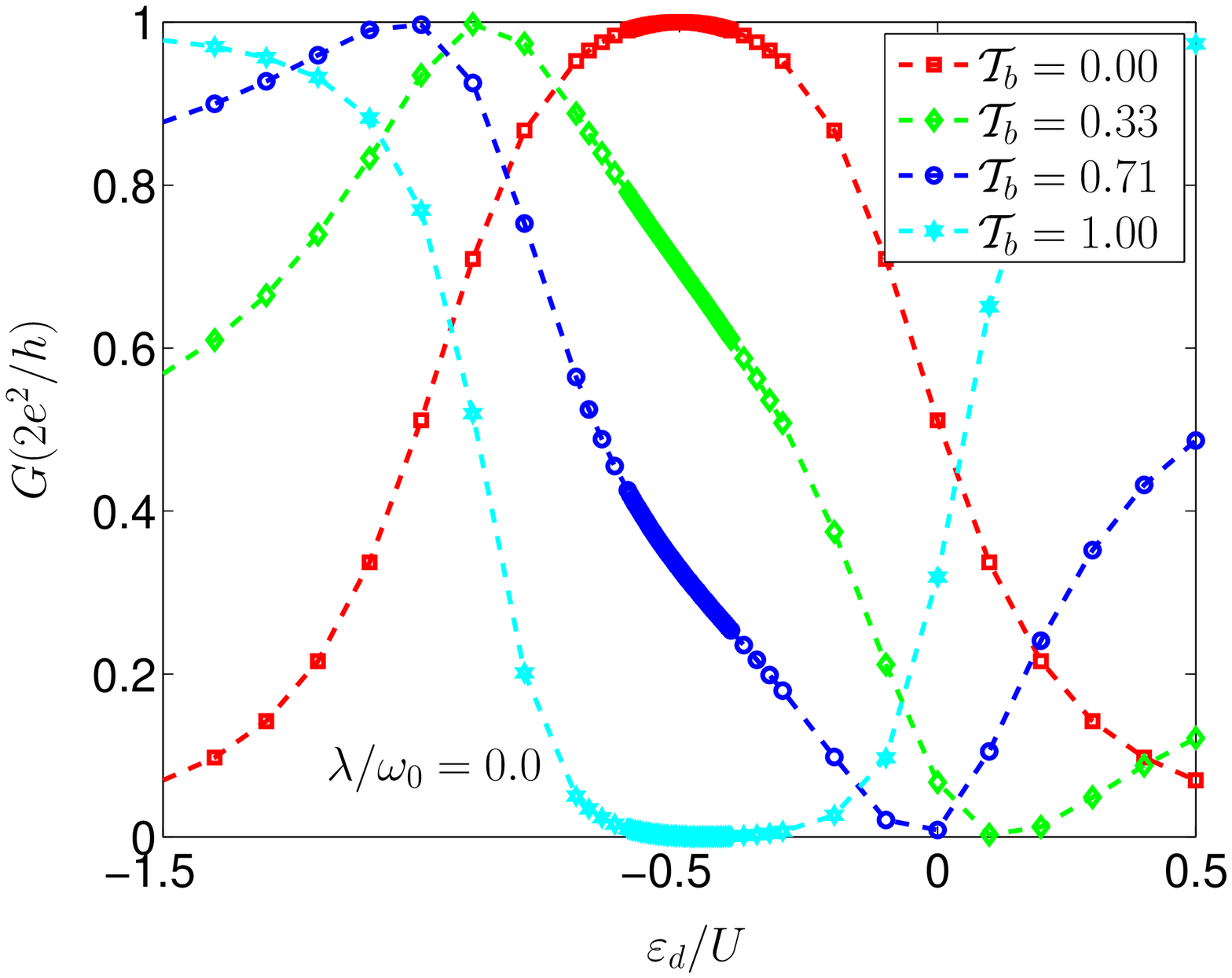}
 \caption{(Color online) Linear conductance $\mathcal{G}$ as a function of a gate voltage $\veps_d$ for different values of background transmission $\calt_b$ at $k_BT=0$.
Parameters are $U=0.1$, $2\Gamma = 0.016$, $\lambda = 0$, and $\varphi = 0$.}
\label{fig:GL00}
\end{figure}

In the following we analyze the local density of states of the molecule  for
$\varphi=0$ both in the weak and strong e-ph coupling.
We start by showing in Fig.~\ref{dos1} the DOS in the absence of the direct
channel, i.e., $W=0$ for the symmetric case $\veps_d=-U/2$. Here, the main effect of the e-ph interaction
is to broad the Kondo resonance when $\lambda/\omega_0<1$  and to shrink it when
$\lambda/\omega_0>1$ (see for example the curve corresponding to
$\lambda/\omega_0=1.2$  which corresponds to the strong e-ph coupling regime).
In this case, $T_K$ becomes very small as expected from
Eq. (\ref{eq:tkstrong}).
When the direct channel is connected, we see in Fig.~\ref{dos2} that the Kondo resonance
is almost
unaffected in the weak e-ph coupling regime ($\lambda/\omega_0 <1$). However, there
is a remarkable feature, the DOS becomes asymmetric. The asymmetry in the DOS indicates
clearly the electron-hole (e-h) symmetry breaking. This lack of e-h
symmetry has dramatic consequences in the strong e-ph coupling regime
where the Kondo effect is due to
charge fluctuations.
%In the absence of $\calt_b$ we showed in Sec.~II B 2
%that charge Kondo regime is very sensitive to
%deviations from the e-h symmetry point (characterized by
%$\Delta$). When $\Delta$ is of the order of the charge Kondo scale then the
%Kondo resonance is effectively suppressed.
%In the AB geometry the lack of e-h symmetry arises from the direct
%path that breaks this symmetry.  
Fig.~\ref{dos3} shows the behavior of the
DOS when $\calt_b\neq 0$. For $\calt_b=0$ we see that the charge Kondo
resonance is pinned at $E_F$ due to the particle-hole symmetry (remember that
$\veps_d=-U/2)$. As long as $\calt_b$ is turn on, the e-h symmetry
is broken and the charge Kondo effect is rapidly destroyed.
All these features are displayed in the linear conductance. Here, we have analyzed
the linear conductance for different
e-ph couplings, different background transmission
$\calt_b$ and Aharonov-Bohm phases.  Fig.~\ref{fig:GL00} displays the zero temperature
linear conductance $\calG$ as a function of the level position $\veps_d$ in the
absence of the e-ph coupling, i.e., $\lambda=0$. Different curves correspond
to different values of background transmission $\calt_b$. For $\calt_b=0$ the conductance reaches the unitary limit, i.e.,
$\mathcal{G}=2e^2/h$ in the Kondo regime ($-U+\Gamma<\veps_d<-\Gamma$) and then
it  drops
rapidly when charge fluctuations are turned on, i.e., whenever $-\Gamma<\veps_d<\Gamma$
and $-U-\Gamma<\veps_d<-U+\Gamma$. Finally, in the empty orbital regime the
molecule charge $n_d\approx
0, 2$ leading to a vanishing conductance. This scenario is totally modified
when the direct path is connected. In such a case, interference effects
between the resonant and non-resonant paths give rise to asymmetric line-shape
for the conductance and eventually a zero conductance region for $\calt_b=1$ in
the Kondo regime, due to the destructive interference. There is however a tiny
feature in the linear conductance not noticed so far due to the lack of
e-h symmetry. The Kondo plateau (for $\calt_b=1$) is not totally symmetric with
respect to $\veps_d/U=-0.5$ as it should be in the case of e-h
symmetry. This is because the direct path breaks this symmetry. However, 
this has a negligible impact when the Kondo correlations are generated
by spin-fluctuations as we have shown. 
However, the lack of e-h symmetry has stronger consequences on the conductance 
in the strong e-ph coupling corresponding to the charge Kondo effect.
%
%\begin{figure}
%\centering
%\includegraphics[width=0.45\textwidth]{figure6}
%\caption[Linear conductance $G$ at zero temperature versus gate voltage $\veps_d$]
%{Linear conductance $\mathcal{G}$ as a function of a gate voltage $\veps_d$ for different values of background transmission $\calt_b$  at $k_BT=0$.
%Parameters are $U=0.1$, $2\Gamma = 0.016$, $\lambda/\omega_0=0.4$, $\omega_0 = 0.05$, and $\varphi = 0$.}
%\label{fig:GL04}
%\end{figure}
%

Now, let us discuss the conductance behavior with $\calt_b$ in terms of
 resonant and non-resonant phase shifts. In the Kondo regime ($\lambda=0$) we can safely assume that
 the resonant phase shift $\delta_{res} = \pi/2$. Taking this fact into account
Eq.~\eqref{eq:tef} is greatly simplified
\beq\label{eq:cos}
\calt(E_F) [\delta_{res} = \pi/2] = \alpha \left[1 - \calt_b \cos^2(\varphi) \right]\,,
\edq
Therefore the linear conductance decreases as long as we open the direct path
by increasing $\calt_b$. It is important to realize that Kondo correlations
are not destroyed in spite of the fact that linear conductance is
suppressed. The reduction of the conductance is rather due to interference
between the lower and the upper paths. Notice, that here the conductance
reduction is maintained for any value of $\veps_d$ in the Kondo regime ($-U+\Gamma<\veps_d<-\Gamma$), whereas
for a noninteracting quantum dot this is only true for the resonant condition
 when the resonant level matches with the Fermi energy in which the phase is
$\delta_{res}=\pi/2$. Finally, away from the Kondo region
$\delta_{res} \to 0,\pi$, and then $\calt (E_F)$ is equivalent to
the non-resonant transmission probability $\calt_b$. In general for
arbitrary $\delta_{res}$ and $\varphi=0$, we obtain \beq \calt(E_F)
=  \left( \calt_b + \alpha\calr_b\right)
\sin^2(\delta_{res}-\delta_{non})\,. \edq This expression leads to a
conductance peak whenever $\delta_{res} = \delta_{non} \pm \pi/2$
and a vanishing conductance for $\delta_{res} =\delta_{non}$. For
$\calt_b = 0$, in the Kondo region, the conductance reaches its
unitary value ($\delta_{res}=\pi/2$ and $\delta_{non}=0$) and
vanishes outside ($\delta_{res}=\delta_{non}=0$). Now we turn on
$\calt_b$ and the linear conductance is gradually suppressed in the
Kondo regime when $\calt_b$ is augmented [$\delta_{res} =\pi/2$ and
$\cot (\delta_{non}) = \sqrt{\alpha \calr_b/\calt_b}$, i.e.,
$\delta_{non}\to 0$ for $\calt_b\to 0$ and $\delta_{non}\to \pi/2$
for $\calt_b\to 1$]. Outside the Kondo regime, $\calG$ increases
($\delta_{res} =0$ and $\delta_{non}\neq 0$). Eventually, for
$\calt_b=1$,  $\mathcal{G}$ vanishes in the Kondo regime
($\delta_{non,res} = \pi/2$). Away from this regime, the conductance
presents a peak since $\delta_{non} = \pi/2$ and $\delta_{res} = 0$.

\begin{figure}
\centering
\includegraphics[width=0.45\textwidth]{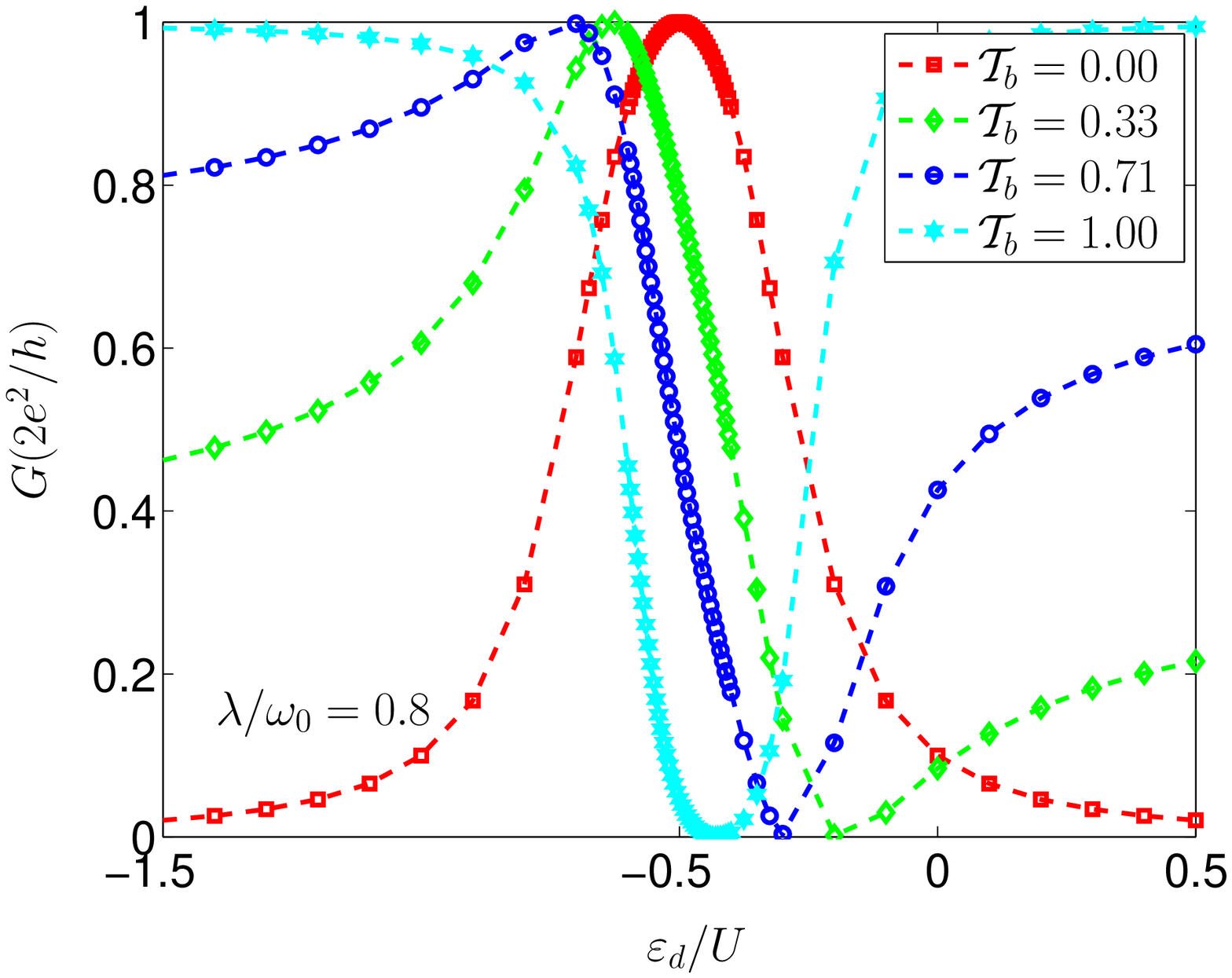}[t]
 \caption{(Color online) Linear conductance $\mathcal{G}$ as a function of a gate voltage
   $\veps_d$ for different values of background transmission $\calt_b$ at
   $k_BT=0$. Parameters are $U=0.1$, $2\Gamma = 0.016$, $\lambda/\omega_0=0.8$, 
   $\omega_0 = 0.05$,and $\varphi = 0$.}
\label{fig:GL08}
\end{figure}
\begin{figure}
\centering
\includegraphics[width=0.45\textwidth]{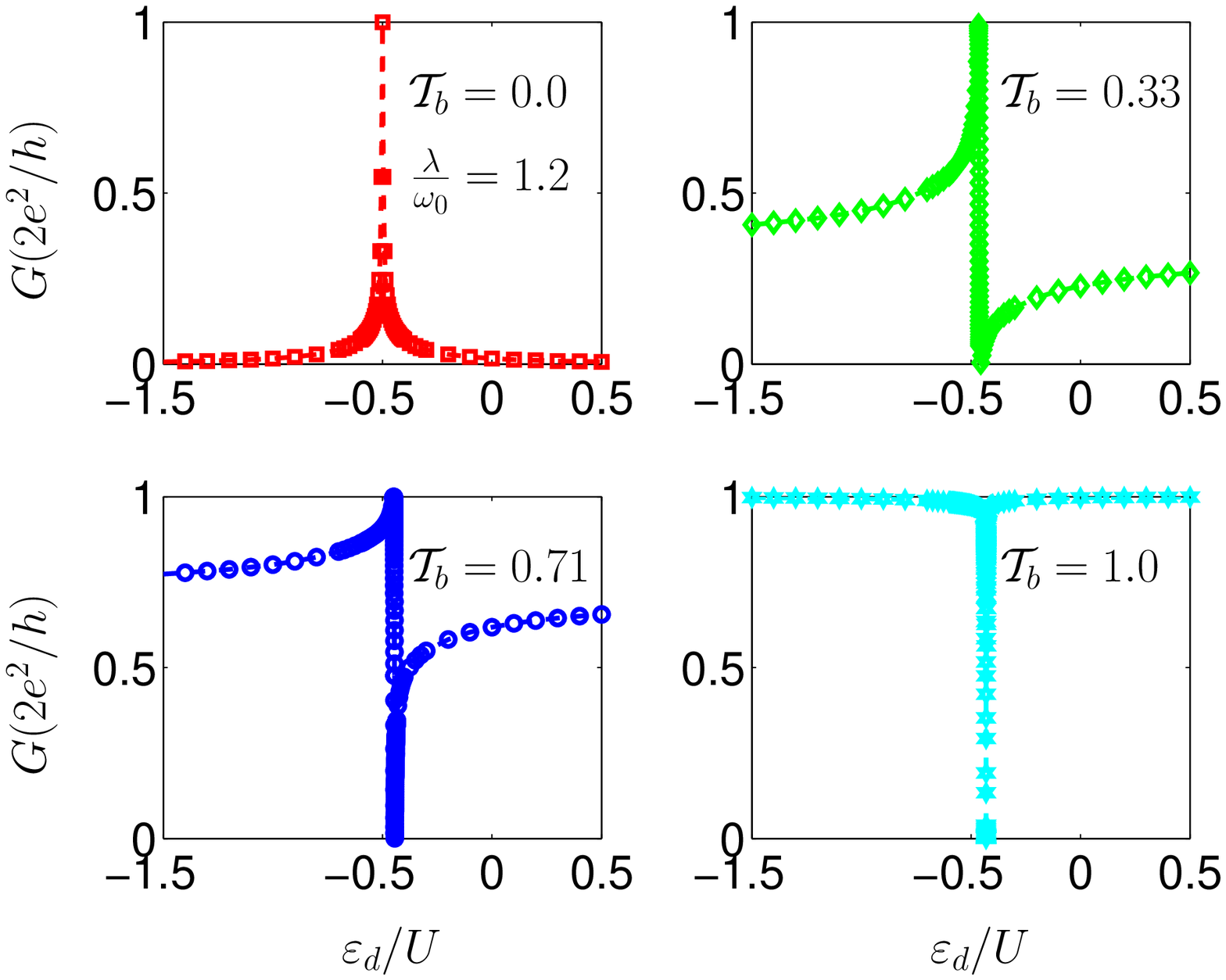}[t]
 \caption{(Color online) Linear conductance $\mathcal{G}$ versus gate voltage
   $\veps_d$ for different values of $\calt_b$ at $k_BT=0$ in the strong
   e-ph coupling regime, $\lambda/\omega_0=1.2$. Parameters are $U=0.1$, $2\Gamma = 0.016$, 
   $\omega_0 = 0.05$, and $\varphi = 0$.}
\label{fig:GL12}
\end{figure}
\begin{figure}
\centering
\includegraphics[width=0.45\textwidth]{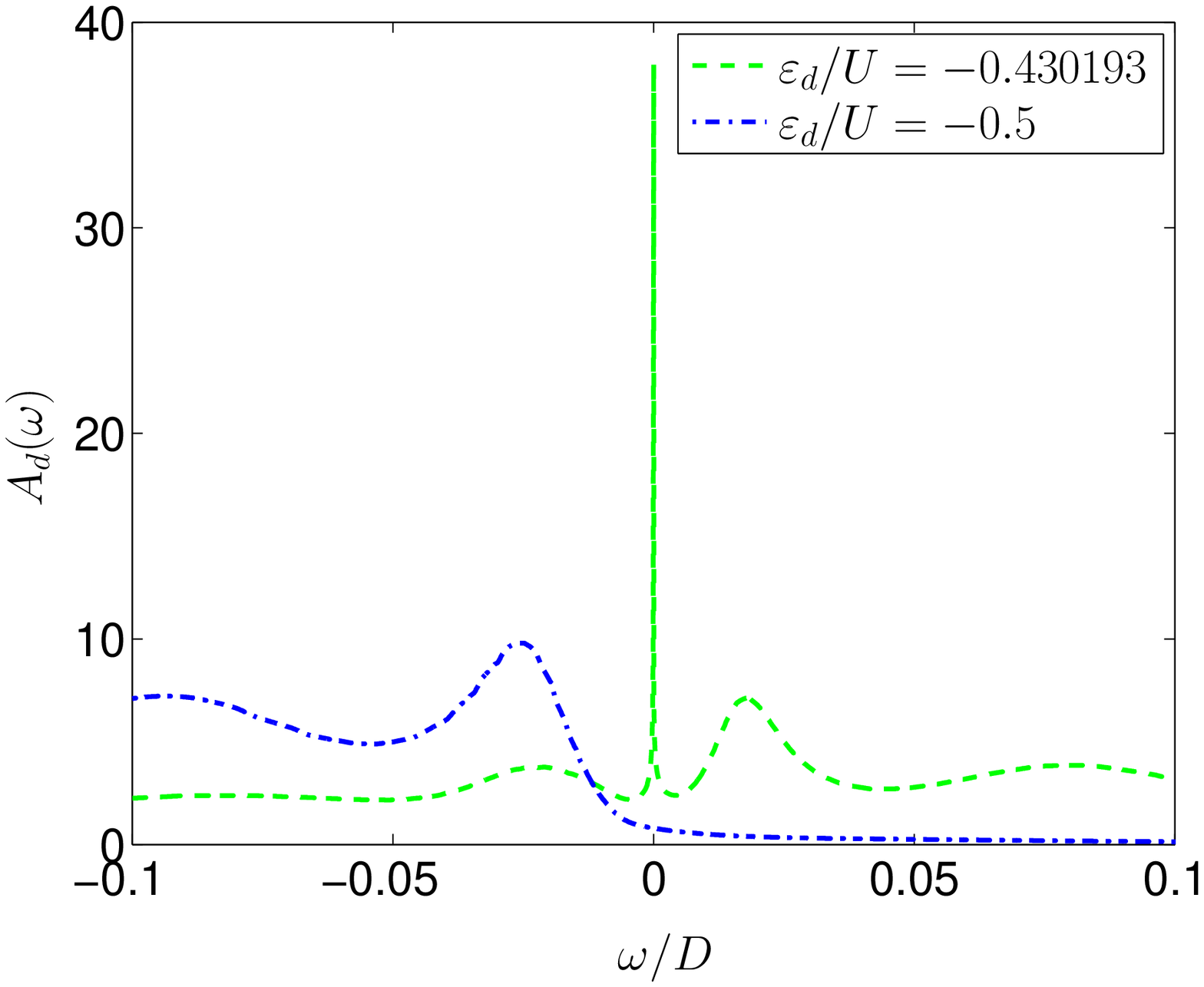}[t]
 \caption[Electronic spectral density $A_d(\omega)$ as a function of $\omega$]
{(Color online) Electronic spectral density $A_d(\omega)$ at two different
  values of $\veps_d$ with $\calt_b=1$. Parameters are $\veps_d = -0.05$, $U=0.1$, $2\Gamma=0.016$, $\varphi = 0$, and $k_BT=0$.}
\label{figrestoration}
\end{figure}

This physical scenario is modified when $\lambda\neq 0$ as shown in
Fig.~\ref{fig:GL08}.  The main difference with the
$\lambda=0$ case (cf.~Fig.~\ref{fig:GL00}) is that
the conductance width decreases due to the presence of
 phonons. This conductance width is roughly given by the e-e
 repulsion, since $U$ is effectively
 renormalized by the electron- phonon interaction ($U_{\rm eff}=U-2\lambda^2/\omega_0$) the Kondo plateau
 region is reduced as long as $\lambda/\omega_0$ increases.  A more
striking feature observed in the linear conductance is the evident
lack of e-h symmetry when $\calt_b\neq 0$. In
Fig.~\ref{fig:GL00} when $\calt_b=1$ the conductance curve is not
symmetric with respect to $\veps_d/U=-0.5$ reflecting the lack of
e-h symmetry. The effect of the e-h symmetry
breaking is more pronounced as long as $\lambda/\omega_0$ grows and the
system enters in the charge Kondo effect originated by the bipolaronic attraction. In this regime, the lack
of e-h symmetry for $\veps_d/U=-0.5$ results in a complete
suppression of the charge Kondo effect ($\lambda/\omega_0>1$) with strong
consequences in the linear conductance. The linear conductance for
the bipolaron regime (strong $e$-$ph$ coupling limit) is plotted in Fig.~\ref{fig:GL12}. In the
absence of the direct path the conductance consists of a very narrow
peak of unitary conductance around $\veps_d/U=-0.5$. The width of
such a peak reflects the fact that the charge Kondo effect occurs
only for $\Delta\approx T_K$ (with $T_K$ being the charge Kondo
temperature). Any deviation from the e-h symmetry point
makes the conductance drops very rapidly to zero. To turn on
$\calt_b$ has two different consequences. First, the direct path
breaks the e-h symmetry and therefore the charge Kondo
effect. However, by tuning properly $\veps_d/U$ the Kondo resonance
can be restored. Second, once the Kondo resonance is restored at
certain value of $\veps_d/U$, the interference between the Kondo
resonance and the direct path can occur. These two effects are
nicely illustrated in Figs.~\ref{fig:GL12} and~\ref{figrestoration}.
Let us consider the linear conductance for $\calt_b=0.33$
(dot-dashed curve in Fig.~\ref{fig:GL12}). Here,  $\calG$ shows a
Fano-like shape where the maximal and minimal conductance take place
for a value of $\veps_d/U$ slightly above the symmetric case
($\veps_d/U=-0.5$). Around this value the DOS displays a charge
Kondo resonance (not shown here). Even more evident is the effect of
the direct path when  $\calt_b=1$. As shown in Fig.~\ref{fig:GL12}
the complete suppression of the linear conductance occurs for
$\veps_d/U=-0.43$ since the Kondo restoration takes place at this
gate value. The restoration of the Kondo resonance is showed in
Fig.~\ref{figrestoration}. Here, the DOS displays a Kondo resonance
at $E_F$. Nevertheless, the restoration of the Kondo effect does not
imply the restoration of the e-h symmetry since the DOS
for $\veps_d/U=-0.43$ is asymmetric (see in
Fig.~\ref{figrestoration}).
\begin{figure}
\centering
\includegraphics[width=0.45\textwidth]{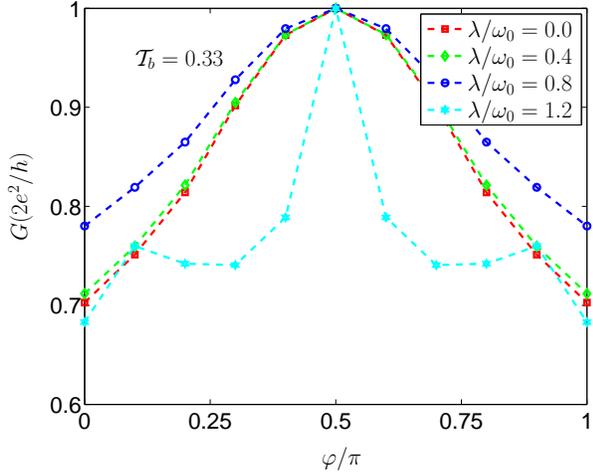}
\caption{(Color online) AB oscillations of the linear conductance $\calG$ at $k_BT=0$ as a
  function of $\varphi$ for $\calt_b=0.33$. Parameters are $\veps_d = -0.05$, $U=0.1$, $2\Gamma = 0.016$, and $\omega_0 = 0.05$.}
\label{ab1}
\end{figure}

\begin{figure}
\centering
\includegraphics[width=0.45\textwidth]{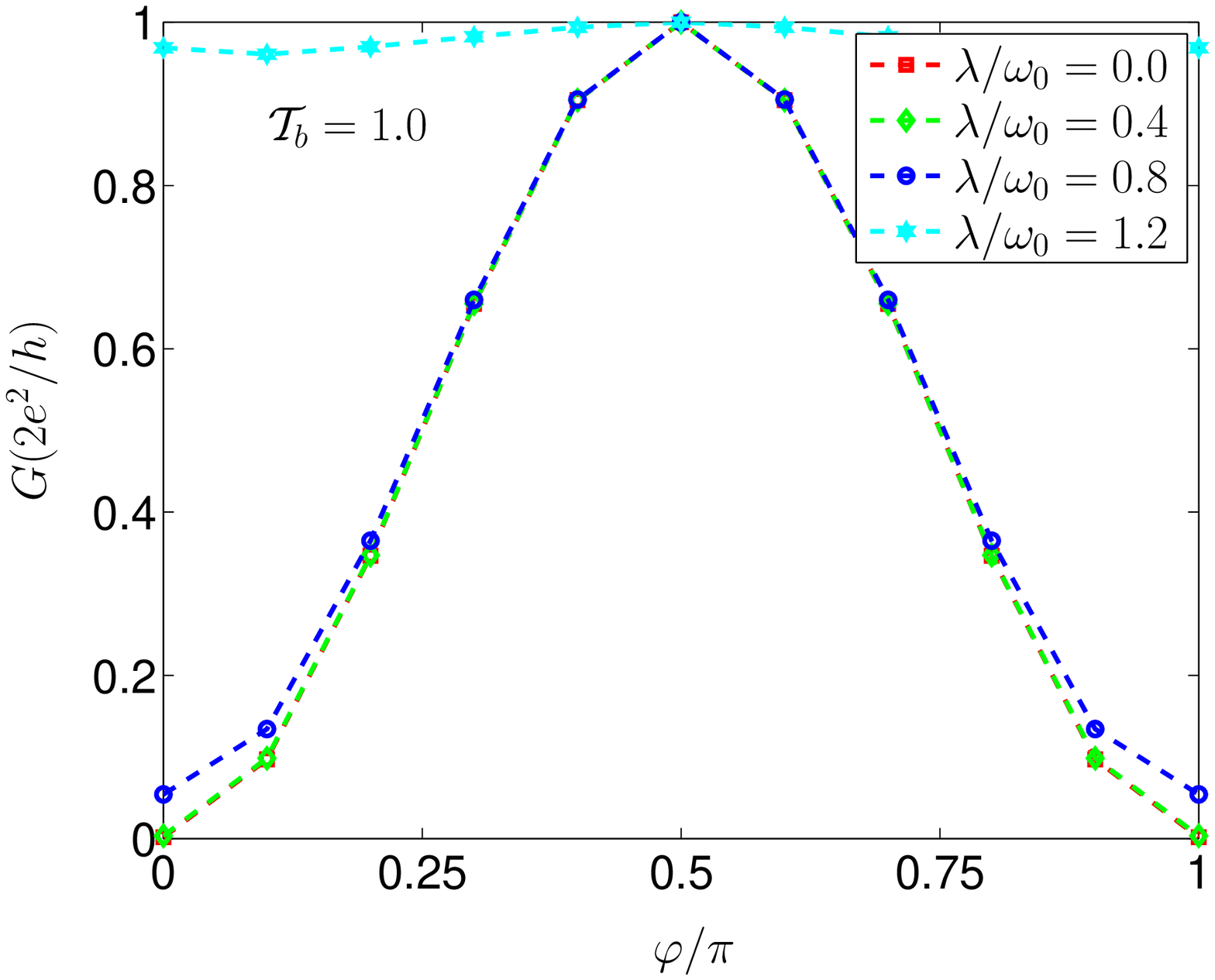}
\caption{(Color online) AB oscillations of the linear conductance $\calG$ at $k_BT=0$ as a
  function of $\varphi$ for $\calt_b=1$.
Parameters are $\veps_d = -0.05$, $U=0.1$, $2\Gamma = 0.016$, and $\omega_0 = 0.05$.}
\label{ab2}
\end{figure}

\begin{figure}
\centering
\includegraphics[width=0.45\textwidth]{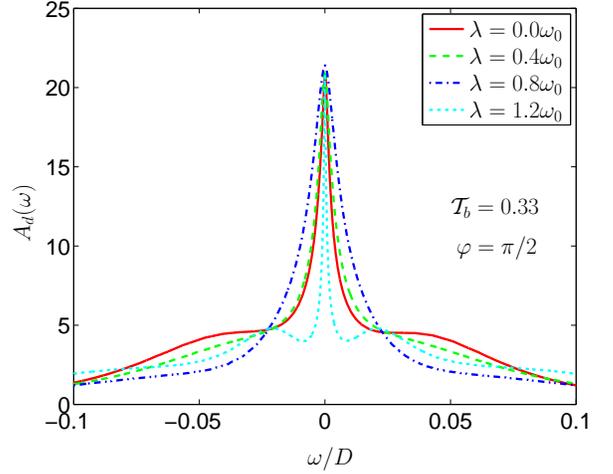}
 \caption[$A_d(\omega)$ as a function of $\omega$]
{(Color online) Electronic spectral density $A_d(\omega)$ for different values of $\lambda/\omega_0$ and $\calt_b=0.33$. Parameters are $\veps_d = -0.05$, $U=0.1$, $2\Gamma=0.016$, $\varphi = \pi/2$, and $k_BT=0$.}
\label{fig2pi}
\end{figure}

Finally we comment the results for the linear conductance as a
function of the AB phase for the two regimes. In Fig.~\ref{ab1} and
Fig.~\ref{ab2}, we display the linear conductance $\mathcal{G}$ as a
function of the AB phase $\varphi$. We have considered only the
symmetric case, $\veps_d = -U/2$. From Eq.~\eqref{eq:tef} we note
that $\calG(\varphi) = \calG(-\varphi)$. Thus, we  plot only
$\varphi \in [0,\pi]$. As we mentioned the direct path breaks the
e-h symmetry. This is true for any value of $\varphi$
except for the special case of $\varphi=\pi/2$. For $\varphi=\pi/2$
the charge Kondo state is not destroyed by the direct channel and
the linear conductance is given by means of Eq.~(\ref{eq:cos}). This
is clearer if we use the parity basis where the odd and even
channels are equally coupled to the molecule and the e-h
symmetry remains unaltered independently of $\calt_b$. To illustrate
this we have plotted in Fig.~\ref{fig2pi} the spectral density at
$\varphi=\pi/2$ for various values of the background transmission
$\mathcal{T}_b$. Having this in mind we can easily explain the
results plotted in Fig.~\ref{ab1}. In the absence of e-ph
coupling, the linear conductance for the Kondo regime follows
Eq.~(\ref{eq:cos}). For the weak e-ph coupling regime
($\lambda/\omega_0=0.4$, and $\lambda/\omega_0=0.8$) Eq.~(\ref{eq:cos}) is still valid
to describe $\calG(\varphi)$ since for $\veps_d=-0.5$ the system is
found in the purely spin Kondo regime, only $U\rightarrow U_{eff}$
and $\veps_d\rightarrow \veps_{d,eff}$. However, for $\lambda/\omega_0=1.2$
the system is in the strong e-ph coupling regime and the Kondo
effect is partially suppressed for $\varphi\neq\pi/2$ due to the
e-h symmetry breaking. For $\varphi\approx \pi/2$ the
conductance shows a peak and only for
 $\varphi=\pi/2$ the conductance is unitary due to the presence of the
charge Kondo resonance and its constructive interference with the
non-resonant path. For $\calt_b=0.33$ and $\varphi\approx 0,\pi$ the
Kondo resonance is not totally quenched (see Fig.~\ref{dos3} for
$\varphi=0$) and therefore the conductance is not only reduced to
the background transmission but it has some contributions coming
from interference between the Kondo resonant path and the direct
path. 

Fig.~\ref{ab2}
shows $\calG$ versus $\varphi$ when the direct path is completely
opened, i.e., $\calt_b=1$. For $\lambda/\omega_0=0,0.4,0.8$ all these curves
follows Eq.~(\ref{eq:cos}) that results from the interference
between the spin Kondo resonance and the direct channel. However for
$\lambda/\omega_0=1.2$ the conductance shows only contribution of the
background transmission probability and a tiny contribution
attributed to the interference between a strong suppressed Kondo
effect and the direct path that makes not unitary the conductance
for $\varphi\approx \pi,0$. At $\varphi=\pi/2$ the conductance
reaches the unitary value since for this special phase value the
charge Kondo effect is not destroyed and we have constructive
interference.

\section{Conclusions}\label{conclusions}
 We have analyzed the linear transport properties of an AB
 interferometer with a molecule inserted in one of its arms. The vibronic
 internal degrees of freedom of the molecule are described by means of phonons
 modes coupled to the molecular occupation. 
Depending on the e-ph coupling we distinguish two different scenarios, the weak e-ph coupling
 limit where the electronic transport occurs as single electron
 tunneling events and the strong e-ph interaction which leads to bipolaronic attraction 
between electrons. Here the tunneling through
 the molecule takes place in pair of electrons. We investigate the low
 temperature limit where Kondo correlations emerge when the molecule is
 strongly coupled to the leads. Depending on the
e-ph coupling the Kondo
 correlations have a different origin. For the weak e-ph coupling
 the Kondo effect corresponds to the spin Kondo effect, while for a strong
 e-ph interaction the bipolaron formation leads to the charge Kondo effect
 with fluctuations between zero and double charge states (bipolaron state). 
The effect of the direct path breaks the
 e-h symmetry with dramatic consequences in the transport properties.
 We show that the direct path destroys completely the charge Kondo effect and
 suppresses the interference between the resonant and non-resonant
 paths. By a proper tuning of the molecular level position, it is possible to
 restore the Kondo resonance but not the e-h symmetry. In this case,
 when the charge Kondo state is revived, interference effects come up.  The usual 
spin Kondo effect is less sensitive to the
 e-h symmetry and is therefore less affected  by the direct path. In
 this case, the transport properties in the weak e-ph coupling regime are
 dominated by the interferences between the spin Kondo effect and the direct path.
In order to obtain these results, we have chosen values for the parameters 
which are in the range of typical experiments in molecular transport and we are
therefore confident that our predictions can be tested in the near future.

\begin{acknowledgments}
We acknowledge D. S\'anchez and R.~\v{Z}itko for their fruitful
  discussions. PS would like to acknowledge O. Entin-Wohlman and
  A. Aharony for an earlier collaboration on related issues. R.L. and J.S.L
  acknowledge financial support from grant  FIS2008-00781 and
  the Conselleria d'Innovaci\'o , Interior i Justicia (before Conselleria d'Economia, Hisenda i Innovaci\'o), Govern  de les Illes
  Baleares. G.P acknowledges finnantial support from grant: MEC: MAT2008-02626.
\end{acknowledgments}
\appendix

\section{Mapping to a Kondo Hamiltonian}\label{app1}
In this appendix we give the details for the derivation of effective
 Hamiltonian using the projection method. This consists of a second order Rayleigh-Schr\"odinger perturbative calculation on the tunneling Hamiltonian, $\calh_T$.\cite{hew93}~We analyze the two cases,
 namely, the weak e-ph coupling and the strong
 e-ph coupling regime.

Let us take $\psi_n$ as the component in which the occupation
number of the molecule is $n_d$, then we can write the following matrix
derived from the Schr\"odinger equation

\beq
\left[ \begin{array}{ccc}
{E - \cal{H}}_{00} & - {\cal{H}}_{01} & - {\cal{H}}_{02} \\
{- \cal{H}}_{10} & E - {\cal{H}}_{11} & - {\cal{H}}_{12} \\
{- \cal{H}}_{20} & - {\cal{H}}_{21} & E - {\cal{H}}_{22}
\end{array}\right]
\left[ \begin{array}{c}
\psi_0 \\
\psi_1 \\
\psi_2
\end{array} \right]
= 0,
\edq
where ${\cal{H}}_{nn'} = P_n {\cal{H}} P_{n'}$.
Here, $P_n$ are the local projectors defined as
\begin{align}
& P_{0} = | 0 \rangle {}\langle 0 | ~\,= ( 1 - n_{\up})(1 - n_{\downarrow})\,, \nn \\
& P_{\up} = | \up \rangle {}\langle \up | = n_{\up}(1 - n_{\downarrow})\,, \nn \\
& P_{\downarrow} = | \downarrow \rangle {}\langle \downarrow | = n_{\downarrow}(1 - n_{\up})\,, \nn \\
& P_{2} = | \uparrow\downarrow \rangle {}\langle \uparrow\downarrow | ~\,= n_{\up}n_{\downarrow},
\label{projector}
\end{align}
where
\beq
P_{0} + P_{\up} + P_{\downarrow} + P_{2} = 1.
\edq
In this fashion the effective Hamiltonian will be given by (we only write the
off-diagonal contributions)
\begin{multline}
\label{effectiveh}
\calh_{eff} = \calh_{1 0}\frac{1}{E_{1,0} - E_{0,m}}\calh_{01}
\\
+  \calh_{12}\frac{1}{E_{1,0} - E_{2,m}}\calh_{21}\,,
\end{multline}
where the energies $E_{1,0}$, $E_{0,m}$, $E_{2,m}$ are the eigenenergies of
the isolated molecule coupled to the local phonon mode.

\subsection{Weak \emph{electron-phonon} coupling limit}
The eigenenergies of the isolated molecule and the local projectors to perform
the second-order perturbative calculation are given respectively by
\bes
\begin{align}\nonumber
E_{1,0} &=  \veps_d\,, \\ \nonumber
E_{0,m} &=  -\frac{\lambda^2}{\omega_0} + m\omega_0\,, \\ \nonumber
E_{2,m} &= -\frac{\lambda^2}{\omega_0} + 2\veps_d + U + m\omega_0\,, \\ \nonumber
\calh_{01} &= \sqrt{2}\sum_{\ell,k,\sm,m} V_{\ell} \langle m | \calu^{-} | 0\rangle c_{\ell k\sm}^{\dag} d_{\sm}(1-n_{d\bsm})\,,  \\ \nonumber
\calh_{21} &= \sqrt{2}\sum_{\ell,k,\sm,m} V_{\ell}^{\ast} \langle m | \calu^{+} | 0\rangle \text{sgn}(\sm) d_{\sm}^{\dag} n_{d\bsm} c_{\ell k \sm} \,,
\end{align}
\eds
with $\bsm = -\sm$. The tunneling amplitudes for the even and odd channels are respectively,
$V_{e} = \cos(\varphi/2)V$ and $V_{o} = \sin(\varphi/2)V$. Using the above expressions the effective Hamiltonian is
\begin{multline}
\calh_{10}\frac{1}{E_{1,0} - E_{0,m}} \calh_{01}
= 2\sum_{\alpha,p,\sm} \sum_{\beta,q,\sm'} \sum_m \frac{V_{\alpha}^{\ast} V_{\beta}|\langle m|\calu^{-}|0\rangle|^2}{\frac{\lambda^2}{\omega_0}+\veps_d-m\omega_0}\\
(1-n_{d\bsm})d_{\sm}^{\dag}c_{\alpha p\sm}c_{\beta q
  \sm'}^{\dag}d_{\sm'}(1-n_{d\bsm'})\,,
\\
\calh_{12}\frac{1}{E_{1,0} - E_{2,m}} \calh_{21}
= 2\sum_{\alpha,p,\sm} \sum_{\beta,q,\sm'} \sum_m \frac{V_{\alpha} V_{\beta}^{\ast}|\langle m|\calu^{+}|0\rangle|^2}{\frac{\lambda^2}{\omega_0}-\veps_d-U-m\omega_0}\\\
\sgn(\sm)\sgn(\sm') c_{\alpha p\sm}^{\dag}n_{d\bsm}d_{\sm}d_{\sm'}^{\dag}n_{d\bsm'}c_{\beta q \sm'}\,,
\end{multline}
Summing over the spin indices all these contributions yields
\begin{align*}
& -\sum_{\alpha,\beta, p,q} J_{0\alpha\beta} (1-n_{d\down})d_{\up}^{\dag}d_{\up}(1-n_{d\down}) c_{\alpha p\up}c_{\beta q\up}^{\dag} \,,
\\
&- \sum_{\alpha,\beta,p,q} J_{2\alpha\beta} n_{d\down}d_{\up}d_{\up}^{\dag} n_{d\down} c_{\alpha p\up}^{\dag}c_{\beta q\up}\,,
\qquad \sm = \up, \sm' = \up, \\
& -\sum_{\alpha,\beta, p,q} J_{0\alpha\beta} (1-n_{d\down})d_{\up}^{\dag}d_{\down}(1-n_{d\up}) c_{\alpha p\up}c_{\beta q\down}^{\dag} \,
\\
&- \sum_{\alpha,\beta,p,q} J_{2\alpha\beta} n_{d\down}d_{\up}d_{\down}^{\dag} n_{d\up} c_{\alpha p\up}^{\dag}c_{\beta q\down}\,,
\qquad \sm = \up, \sm' = \down, \\
& -\sum_{\alpha,\beta, p,q} J_{0\alpha\beta} (1-n_{d\up})d_{\down}^{\dag}d_{\up}(1-n_{d\down}) c_{\alpha p\down}c_{\beta q\up}^{\dag}
\\
&- \sum_{\alpha,\beta,p,q} J_{2\alpha\beta} n_{d\up}d_{\down}d_{\up}^{\dag} n_{d\down} c_{\alpha p\down}^{\dag}c_{\beta q\up}\,,
\qquad \sm = \down, \sm' = \up ,\\
& -\sum_{\alpha,\beta, p,q} J_{0\alpha\beta} (1-n_{d\up})d_{\down}^{\dag}d_{\down}(1-n_{d\up}) c_{\alpha p\down}c_{\beta q\down}^{\dag}
\\
&- \sum_{\alpha,\beta,p,q} J_{2\alpha\beta} n_{d\up}d_{\down}d_{\down}^{\dag} n_{d\up} c_{\alpha p\down}^{\dag}c_{\beta q\down}\,,
\qquad \sm = \down, \sm' = \down\,,
\end{align*}
where we have defined the following quantities as the exchanged couplings
\beq\label{eq:jalbe}
J_{0\alpha\beta} = -\sum_m \frac{2V_{\alpha}^{\ast} V_{\beta}|\langle
  m|\calu^{-}|0\rangle|^2}{\frac{\lambda^2}{\omega_0}+\veps_d-m\omega_0},
\end{equation}
\beq
J_{2\alpha\beta} = -\sum_m \frac{2V_{\alpha} V_{\beta}^{\ast}|\langle m|\calu^{+}|0\rangle|^2}{\frac{\lambda^2}{\omega_0}-\veps_d-U-m\omega_0}\,.
\label{eq:j2albe}
\end{equation}
Using these results Eq.~(\ref{effectiveh}) reads
\begin{multline}
\calh_{eff}=\sum_{\alpha,\beta,p,q} J_{0\alpha\beta} \bigg[ n_{d\up}(1-n_{d\down}) (1 - \delta_{\alpha,\beta}\delta_{p,q}) c_{\beta q\up}^{\dag}c_{\alpha p\up}
 \\+ d_{\up}^{\dag}d_{\down} c_{\beta q\down}^{\dag}c_{\alpha p\up}
+ d_{\down}^{\dag}d_{\up} c_{\beta q\up}^{\dag}c_{\alpha p\down}
 \\+ n_{d\down}(1-n_{d\up}) (1 - \delta_{\alpha,\beta}\delta_{p,q})c_{\beta q\down}^{\dag}c_{\alpha p\down} \bigg] \\
+ \sum_{\alpha,\beta,p,q} J_{2\alpha\beta} \bigg[ (n_{d\up}-1) n_{d\down} c_{\alpha p\up}^{\dag}c_{\beta q\up}
+ d_{\down}^{\dag}d_{\up} c_{\alpha p\up}^{\dag}c_{\beta q\down}
 \\ + d_{\up}^{\dag}d_{\down} c_{\alpha p\down}^{\dag}c_{\beta q\up}
+ (n_{\down}-1)n_{d\up}c_{\alpha p\down}^{\dag}c_{\beta q\down} \bigg]\,, \\
\label{eq:swint}
\end{multline}

Since the possibility of $p=q$ is extremely small, we can safely drop the
Kronecker delta terms. By employing  the following identities
\bes
\begin{align}
S^+ &= d_{\up}^{\dag}d_{\down}\,,\,\,S^- = d_{\down}^{\dag}d_{\up} \,, \\
S^z &= \frac{1}{2}(d_{\up}^{\dag}d_{\up} - d_{\down}^{\dag}d_{\down})
 = \frac{1}{2}\left[n_{d\up}(1-n_{d\down}) - n_{d\down}(1-n_{d\up})\right]\,, \\
1   &= n_{d\up} + n_{d\down} = n_{d\up}(1-n_{d\down}) + n_{d\down}(1-n_{d\up})\,.
\end{align}
\eds
Equation~\eqref{eq:swint} can be then written as
\begin{multline}\label{heffective}
\calh_{eff}=\sum_{\alpha,\beta,p,q} J^{\alpha\beta} \Biggr[
S^z \cdot \left( c_{\alpha p\up}^{\dag}c_{\beta q\up} - c_{\alpha p\down}^{\dag}c_{\beta q\down}\right)
\\
+ S^+ \cdot  c_{\alpha p\down}^{\dag}c_{\beta q\up}
+ S^- \cdot  c_{\alpha p\up}^{\dag}c_{\beta q\down}
\\
+ \frac{1}{2} \sum_{\alpha,\beta,p,q,\sm} K^{\alpha\beta}  c_{\alpha p\sm}^{\dag} c_{\beta q\sm}\Biggr]\,.
\end{multline}
where $J^{\alpha\beta}=\left(J_{0\alpha\beta} + J_{2\alpha\beta}\right)$ and
$K^{\alpha\beta}=\left(J_{0\alpha\beta} - J_{2\alpha\beta}\right)$
\subsection{Strong \emph{electron-phonon} coupling: the bipolaronic scenario}
Under the particle-hole transformation, Eq.~\eqref{eq:model} becomes
\bes
\begin{align}
\calh_C &= \sum_{\ell=1/2,k,\sm} \veps_{k\sm} \wtc_{\ell k\sm}^{\dag} \wtc_{\ell k\sm}\,
\\
&+ \sum_{k,k',\sm} \left[\sgn(\sm) W e^{i\varphi} \wtc_{2k'\sm}^{\dag} \wtc_{1k\sm} + h.c. \right]\,,\\
\calh_M &= \sum_{\sm} \wteps_{d\sm} \wtd_{\sm}^{\dag} \wtd_{\sm} + \wtU \wtn_{d\up}\wtn_{d\down}\,,
 \\
\calh_T &= \sum_{\ell=1/2,k,\sm} \left( \wV_{\ell k\sm} \wtc_{\ell k\sm}^{\dag} \wtd_{\sm} + h.c.\right)\,, \\
\calh_{e-ph} &= \lambda (a+a^{\dag})\left(\wtn_{d\up} - \wtn_{d\down}\right) \,,\\
\calh_{ph} &= \omega_o a^{\dag}a\,,
\end{align}
\label{eq:stmodel}
\eds
where $\wteps_{d\sm}
= U/2 + \sgn(\sm) \Delta /2$, $\wtU = - U$ and
$\Delta = 2\veps_d + U$. Assumming the symmetry condition for the conduction
band  $\veps_{k\sm} = -\veps_{\wtk \sm}$, the tunneling
amplitudes become
\begin{eqnarray}
&\wV_{1k\up} = V_{Lk\up}, \,\, \wV_{1k\down} = V_{R\wtk\down}^{\ast}, \\
&\wV_{2k\up} = V_{Rk\up}, \,\,\wV_{2k\down} = V_{L\wtk\down}^{\ast}\,.
\end{eqnarray}
Now we perform the Lang-Firsov transformation.~\cite{Mahanbook,Lang63} The molecular part
in the presence of phonons ($\calh_M+ \calh_{e-ph}+\calh_{ph}$) in Eq.~\eqref{eq:stmodel} reads
\begin{eqnarray}\label{hamstrong1}
\calh_M &=& \sum_{\sm} \left(\wteps_{d\sm}-\frac{\lambda^2}{\omega_0}\right)
\wtd_{\sm}^{\dag} \wtd_{\sm} \\
\nonumber
&+& \left(\wtU + \frac{2\lambda^2}{\omega_0}\right) \wtn_{d\up}\wtn_{d\down}
+ \omega_0 a^{\dag}a\,,
\end{eqnarray}
and the tunneling part $\calh_T$ is
\begin{eqnarray}\label{hamstrong2}
\calh_T &=& \sum_{\ell=1/2,k,\sm} \left(\wV_{\ell k\sm} e^{-\sgn(\sm)\lambda(a^{\dag}-a)/\omega_0} \wtc_{\ell k\sm}^{\dag} \wtd_{\sm}+ h.c.\right)\,.
\end{eqnarray}
Next we perform a canonical transformation of the  total Hamiltonian
($\mathcal{\tilde{H}}\rightarrow\mathcal{O}^\dagger \mathcal{H}\mathcal{O}$) using the following unitary operator
\beq
\mathcal{O}=e^{\frac{\lambda}{\omega_0}(a^{\dag}-a)(\wtn_{d\up}-\wtn_{d\down})}\,.
\edq
This transformation yields the following eigenstates and eigenvalues for the
isolated molecule (in the equivalent model):
\bes
\begin{align}
|\widetilde{0},m\rangle &= |\widetilde{0}\rangle|m\rangle,
~~~~E_{\widetilde{0},m} = \frac{\lambda^2}{\omega_0} + m\omega_0 ,\\
|\widetilde{\sm},m\rangle &= \calu^{\sgn(\widetilde{\sm})}|\widetilde{\sm}\rangle|m\rangle,
~E_{\widetilde{\sm},m} = \wteps_{d\sm} + m\omega_0, \\
|\widetilde{2},m\rangle &= |\widetilde{2}\rangle |m\rangle,~~
E_{\widetilde{2},m} = \frac{\lambda^2}{\omega_0} + \wteps_{d\sm}   + \wteps_{d\bar\sm} + \wtU + m\omega_0,
\end{align}
\eds
with
\beq
\calu^{\sgn(\widetilde{\sm})} = \exp\left[\sgn(\widetilde{\sm}) \frac{\lambda}{\omega_0}(a^{\dag}-a)\right]\,.
\edq
This define the projectors for the strong e-ph coupling
regime as
\bes
\begin{align}
\calh_{01} &= \sqrt{2}\sum_{\ell,k,\sm,m} \wV_{\ell k\sm} \langle m | \calu^{-\sgn(\sm)} | 0\rangle \wtc_{\ell k\sm}^{\dag} \wtd_{\sm}(1-\wtn_{d\bsm}) \,, \\
\calh_{21} &= \sqrt{2}\sum_{\ell,k,\sm,m} \wV_{\ell k\sm}^{\ast} \langle m | \calu^{+\sgn(\sm)} | 0\rangle \text{sgn}(\sm) \wtd_{\sm}^{\dag} \wtn_{d\bsm} \wtc_{\ell k \sm}\,,
\end{align}
\eds
with $\bsm = -\sm$, and $\wV_{ek\sm} = \cos(\varphi/2)\wV_{k\sm}$ and
$\wV_{ok\sm} = \sin(\varphi/2)\wV_{k\sm}$. With these ingredients the
effective Hamiltonian is given by the sum of the following terms
\bes
\begin{multline}
  \calh_{10}\frac{1}{E_{\sigma,0} - E_{0,m}}\calh_{01} \\
= 2\sum_{\alpha,p,\sm}\sum_{\beta,q,\sm'}\sum_m
\frac{\wV_{\alpha p\sm}^{\ast}\wV_{\beta q\sm'}\langle 0|\calu^{\sgn(\sm)}|m\rangle \langle m|\calu^{-\sgn(\sm')}|0\rangle}
{(E_{\sm,0}+E_{\sm',0})/2 - \frac{\lambda^2}{\omega_0}-m\omega_0}\,,
\\
(1-\wtn_{d\bsm})\wtd_{\sm}^{\dag}\wtc_{\alpha p\sm}\wtc_{\beta q\sm'}^{\dag}\wtd_{\sm'}^{\dag}(1-\wtn_{d\bsm'})
\end{multline}
\begin{multline}
\calh_{12}\frac{1}{E_{\sigma,0} - E_{2,m}}\calh_{21} \\
= 2\sum_{\alpha,p,\sm}\sum_{\beta,q,\sm'}\sum_m
\frac{\wV_{\alpha p\sm}\wV_{\beta q\sm'}^{\ast}\langle 0|\calu^{-\sgn(\sm)}|m\rangle \langle m|\calu^{\sgn(\sm')}|0\rangle}
{(E_{\sm,0}+E_{\sm',0})/2 - \frac{\lambda^2}{\omega_0}-\wteps_{d\sm}-\wteps_{d\bsm} - \wtU - m\omega_0}\,,
 \\
\sgn(\sm)\sgn(\sm')
\wtc_{\alpha p\sm}^{\dag} \wtn_{d\bsm}\wtd_{\sm}\wtd_{\sm'}^{\dag}\wtn_{d\bsm'}\wtc_{\beta q\sm'}\,.
\end{multline}
\eds
The energies $E_{\sm,0}$ of the initial and final states can be different. We eliminate this ambiguity by choosing the average between
 both energy states.
Finally, a similar calculation than for the weak regime, produces the
following exchange couplings
\bes
\begin{align}
&J_{\parallel}^{\alpha\beta} = \frac{8}{U} \sum_m \frac{V_{\alpha} V_{\beta}|\langle m|\calu^{-}|0\rangle|^2}{\frac{2\lambda^2}{\omega_0 U}-1+\frac{2m\omega_0}{U}}\,, \\
&J_{\perp}^{\alpha\beta} = \frac{8}{U} \sum_m
\frac{V_{\alpha} V_{\beta} \langle 0|\calu^{-}|m\rangle\langle m|\calu^{-}|0\rangle}{\frac{2\lambda^2}{\omega_0 U}-1+\frac{2m\omega_0}{U}}\,,
\\
&= \frac{8}{U} \sum_m (-1)^m
\frac{V_{\alpha} V_{\beta} |\langle
  m|\calu^{-}|0\rangle|^2}{\frac{2\lambda^2}{\omega_0
    U}-1+\frac{2m\omega_0}{U}}\,, \nonumber
\end{align}
\eds
Now we can perform an expansion in the large $\lambda/\omega_0^2$ limit
\begin{align*}
&J_{\parallel}^{\alpha\beta} \left[ \frac{8V_{\alpha}V_{\beta}}{U} \right]^{-1} = \frac{e^{-(\lambda/\omega_0)^2}}{\frac{2\lambda^2}{\omega_0 U}-1}
+ \frac{e^{-(\lambda/\omega_0)^2}\left(\frac{\lambda}{\omega_0}\right)^2}{\frac{2\lambda^2}{\omega_0 U}-1+\frac{2\omega_0}{U}} + \cdots \\
&\approx \frac{\omega_0 U}{2\lambda^2} e^{-(\lambda/\omega_0)^2}
\left[ 1+\left(\frac{\lambda}{\omega_0}\right)^2 +\frac{1}{2!}\left(\frac{\lambda}{\omega_0}\right)^4 + \cdots \right] \\
&= \frac{\omega_0 U}{2\lambda^2}\,, \\
&J_{\perp}^{\alpha\beta} \left[ \frac{8V_{\alpha}V_{\beta}}{U} \right]^{-1} = \frac{e^{-(\lambda/\omega_0)^2}}{\frac{2\lambda^2}{\omega_0 U}-1}
- \frac{e^{-(\lambda/\omega_0)^2}\left(\frac{\lambda}{\omega_0}\right)^2}{\frac{2\lambda^2}{\omega_0 U}-1+\frac{2\omega_0}{U}} + \cdots \\
&\approx \frac{\omega_0 U}{2\lambda^2} e^{-(\lambda/\omega_0)^2}
\left[ 1-\left(\frac{\lambda}{\omega_0}\right)^2 +\frac{1}{2!}\left(\frac{\lambda}{\omega_0}\right)^4 + \cdots \right] \\
&= \frac{\omega_0 U}{2\lambda^2} e^{-2(\lambda/\omega_0)^2}\,,
\end{align*}
and finally we note that $J_{\perp}$ is exponentially suppressed in comparison
with $J_{\parallel}$ giving a very small Kondo temperature in the strong
e-ph coupling regime [see Eq.~(\ref{tkstrongdef})].

\section{Electronic and phonon Green functions}
\label{app2}
In this appendix we calculate the electronic and phonon Green functions in
order to calculate the density of states of electrons and the
linear conductance. To this end we employ the
\emph{equation of motion} (EOM) method.\cite{lan66}~In this manner the equation of
motion for the correlator $<A,B>$ in the frequency domain ($z$) is given by
\beq
z\langle\langle A,B \rangle\rangle_{z} + \langle\langle [\calh,A]_-,B \rangle\rangle = \langle [A,B]_{\zeta} \rangle\,,
\label{eq:ceom}
\edq
with $\zeta=+$ if both $A$ and $B$ are fermionic operators, while $\zeta=-$
otherwise. Notice that we have defined as
\bes
\begin{align}
\langle\langle A,B \rangle\rangle_z &= \int_{-\infty}^{\infty} dt~ e^{izt} \langle\langle A,B \rangle\rangle_t\,, \\
\langle\langle A,B \rangle\rangle_t &= -i\Theta(t) \langle [A(t),B]_{\zeta} \rangle\,.
\end{align}
\eds
Here, $\Theta$ denotes the step function, $\langle\cdot \rangle$ is the
thermodynamic average. Next, we obtain the electron Green function in the presence of the
e-ph interaction. We derived the molecule Green function and also the lead-molecule Green function.
For $A=d_{\sigma}$ and $A=c_{\alpha k\sigma}$ with $B=d_{\sigma}^{\dag}$
the equation of motion for the molecule and lead-molecule Green function are respectively
\begin{eqnarray}\label{eq:gdr1}
&(z-\eps_d) \calg_{d\sigma,d\sigma}(z) = 1 + U \langle\langle d_{\sigma}d_{\bar{\sigma}}^{\dag}d_{\bar{\sigma}},d_{\sigma}^{\dag} \rangle\rangle_z
\\+&\lambda \langle\langle d_{\sigma}(a+a^{\dag}),d_{\sigma}^{\dag} \rangle\rangle_z
+ \sum_{\alpha} \sum_k V_{\alpha}^{\ast} \langle\langle c_{\alpha k\sigma},d_{\sigma}^{\dag} \rangle\rangle_z\,,  \nonumber
\end{eqnarray}
and
\begin{eqnarray}
&&(z-\veps_{k\sigma}) \langle\langle c_{\alpha k\sigma}, d_{\sigma}^{\dag}
  \rangle\rangle_z = V_{\alpha} \langle\langle d_{\sigma},d_{\sigma}^{\dag}
  \rangle\rangle_z \nonumber
\\ & + & W e^{i\sgn{(\alpha)} \varphi} \sum_q \langle\langle c_{\bar{\alpha} q\sigma},d_{\sigma}^{\dag} \rangle\rangle_z\,, \label{eq:gdr2}
\end{eqnarray}
where $\sgn{(\alpha)} = -(+)$ for $\alpha = L(R)$ and $\bar{\alpha} = R(L)$.
Here, Eq.~\eqref{eq:gdr2} can be rewritten as
\bes
\begin{align}
\sum_p \calg_{Lp\sigma,d\sigma}^r(z) &= \frac{ \calg_{d\sigma,d\sigma}^r(z)}{1+\xi} \left[ -i\pi\rho_L V_L - \pi^2 \rho_L \rho_R W e^{-i\varphi} V_R \right] \label{eq:gdrr1} \,,\\
\sum_q \calg_{Rq\sigma,d\sigma}^r(z) &= \frac{\calg_{d\sigma,d\sigma}^r(z)}{1+\xi} \left[ -i\pi\rho_R V_R - \pi^2 \rho_L \rho_R W e^{+i\varphi} V_L \right]  \label{eq:gdrr2}\,,
\end{align}
\eds
where $\langle\langle c_{\alpha l \sigma},d_{\sigma}^{\dag} \rangle\rangle_z = \calg_{\alpha l\sigma,d\sigma}^r(z)$.
Taking Eqs.~\eqref{eq:gdrr1}, \eqref{eq:gdrr2}, Eq.~\eqref{eq:gdr1},
and Eq.~\eqref{eq:gdr2} the molecule Green function reads
\begin{eqnarray}
\calg_{d\sigma,d\sigma}^{-1}(z) &=& \scrg_{d\sigma,d\sigma}^{-1}(z)
- U\frac{\langle\langle d_{\sigma}d_{\bar{\sigma}}^{\dag}d_{\bar{\sigma}},d_{\sigma}^{\dag} \rangle\rangle_z}{\calg_{d\sigma,d\sigma}(z)}
\\&-&\lambda \frac{\langle\langle d_{\sigma}(a+a^{\dag}),d_{\sigma}^{\dag} \rangle\rangle_z}{\calg_{d\sigma,d\sigma}(z)}\nonumber\,,
\end{eqnarray}
where
\beq
\scrg_{d\sigma,d\sigma}^{-1}(z) = z - \veps_d - \frac{1}{1+\xi} \left[ -2i\Gamma - 2\pi^2 \rho_0^2 |V|^2 W \cos(\varphi) \right]\,.
\edq
Finally,
$\langle\langle
d_{\sigma}d_{\bar{\sigma}}^{\dag}d_{\bar{\sigma}},d_{\sigma}^{\dag}
\rangle\rangle_z$ and $\langle\langle d_{\sigma}(a+a^{\dag}),d_{\sigma}^{\dag}
\rangle\rangle_z$ are determined from the NRG energy spectrum. In order to do
this we write them using the spectral representation
\begin{multline}
\langle\langle d_{\sm}d_{\bsm}^{\dag}d_{\bsm},d_{\sm}^{\dag} \rangle\rangle_z \\
= \frac{1}{\calz} \sum_{n,m}
\langle n|d_{\sm}^{\dag}|m\rangle\langle m|
d_{\sm}d_{\bsm}^{\dag}d_{\bsm}|n\rangle \left( e^{-\beta E_n} + e^{-\beta E_m}
\right) \\
\left[ \calp \frac{1}{z - (E_n + E_m)} -i \pi \delta(z - E_n + E_m) \right]\,,
\label{eq:d4sp}
\end{multline}
and
\begin{multline}
\langle\langle d_{\sm}(a + a^{\dag}),d_{\sm}^{\dag} \rangle\rangle_z \\
= \frac{1}{\calz} \sum_{n,m}
\langle n|d_{\sm}^{\dag}|m\rangle\langle m| d_{\sm}(a+a^{\dag})|n\rangle
\left( e^{-\beta E_n} + e^{-\beta E_m} \right) \\
\left[ \calp \frac{1}{z - (E_n + E_m)} -i \pi \delta(z - E_n + E_m) \right]\,,
\label{eq:dasp}
\end{multline}
where $\calz$ is the partition function and $ \calp$ denotes the principal
value in the Cauchy sense.

The phonon Green function is calculated in the same manner as the electronic
Green function by using the EOM
method. Now $A=a$ and $B=a^{\dag}$ and then the phonon Green's function reads
\beq
(z-\omega_0) \langle\langle a,a^{\dag} \rangle\rangle_z = 1 + \lambda \langle\langle n_{d},a^{\dag} \rangle\rangle_z\,.
\label{eq:pheom}
\edq
By dividing both sides of Eq.~\eqref{eq:pheom} by $\langle\langle
a,a^{\dag}\rangle\rangle_z$ we get a Dyson type equation for the phonon
propagator
\beq\label{ph2}
\cald^{-1}(z) = \cald_0^{-1}(z) - \lambda \frac{\langle\langle n_{d},a^{\dag} \rangle\rangle_z}{\langle\langle a,a^{\dag} \rangle\rangle_z}\,,
\edq
where $\cald(z) \equiv \langle\langle a,a^{\dag} \rangle\rangle_z$,
$\cald_0^{-1}(z) = z - \omega_0$ and $n_d \equiv \sum_{\sigma} n_{d\sigma} - 1$.
The propagators $\langle\langle
n_{d},a^{\dag} \rangle\rangle_z$ and $\langle\langle
a,a^{\dag}\rangle\rangle_{z}$ are obtained from the NRG spectrum.
Using again the spectral representation we have:
\begin{multline}\label{ph1}
\langle\langle a,a^{\dag}\rangle\rangle_{z} = -\frac{1}{\calz} \sum_{n,m}
|\langle n |a^{\dag}|m\rangle|^2 \left( e^{-\beta E_n} - e^{-\beta E_m}
\right)
\\
\left[\calp \frac{1}{z - (E_n - E_m)} - i\pi \delta\left(z - (E_n -
  E_m)\right) \right]\,,
\end{multline}
and
\begin{multline}
\langle\langle n_d,a^{\dag}\rangle\rangle_{z} = -\frac{1}{\calz} \sum_{n,m}
\langle n |a^{\dag}|m\rangle\langle m| n_d |n\rangle \left( e^{-\beta E_n} -
e^{-\beta E_m} \right)
\\
\left[ \calp \frac{1}{z - (E_n - E_m)} - i\pi \delta\left(z - (E_n -
  E_m)\right) \right]\,.
\label{eq:ndasp}
\end{multline}

This method of calculating the phonon propagator through the EOM technique [see
  Eq.~(\ref{ph2})] is much more accurate than the
direct calculation of this from the NRG spectrum [see
  Eq.~(\ref{ph1})] as it has been shown by Jeon {\it et al.,} in Ref.~[\onlinecite{Jeon03}].   

% We compare the two approaches in Fig.~\ref{phonon1} and  Fig.~\ref{phonon2}. We see that the direct calculation of the phonon  propagator does not develop a narrow resonance around $\omega_0$ whereas the spectral density for the phonon shows this singularity when $\lambda$ enhances.
%\begin{figure}
%\centering
%\includegraphics[width=0.4\textwidth]{phonon_nonrenorm}
%\caption{Direct calculation of $\langle\langle a,a^{\dag}\rangle\rangle$. Refer to Eq.~(\ref{ph1}).}\label{phonon1}
%\end{figure}

%\begin{figure}
%\centering
%\includegraphics[width=0.4\textwidth]{phonon_renormalized}
%\caption{Calculation of $\langle\langle a,a^{\dag}\rangle\rangle$. Refer to Eq.~(\ref{ph2}).}\label{phonon2}
%\end{figure}

\end{document}